\documentclass[a4paper,10pt,twoside]{cpc-hepnp}
\usepackage{CJK,upgreek,fancyhdr}
\usepackage{multicol}
\usepackage{graphicx}
\usepackage{booktabs}
\usepackage{amssymb,bm,mathrsfs,bbm,amscd}
\usepackage[tbtags]{amsmath}
\usepackage{lastpage}

\newcommand{\DDbar}{D\bar{D}}
\newcommand{\psip}{\psi(3686)}
\newcommand{\psipp}{\psi(3770)}
\newcommand{\jpsi}{J/\psi}
\newcommand{\qqbar}{q\bar{q}}
\newcommand{\mbc}{\text{M}_{\text{BC}}}

\usepackage{gensymb}

\usepackage{setspace}

\usepackage[dvipsnames]{xcolor}
\usepackage{hyperref}
\hypersetup{
 pdftitle={},%
 pdfauthor={},%
 pdfsubject={},%
 pdfkeywords={},%
 pdfstartview={},%
 bookmarksopen=true, breaklinks=true, debug=true,
 colorlinks=true, linkcolor=blue, citecolor=red, urlcolor=RoyalBlue,
 hyperfigures=true
}
\def\equationautorefname~#1\null{(#1)\null}

\begin{document}
\begin{CJK*}{UTF8}{gkai}
%% note here we use UTF8 and gbsn(simpe chinese character)
%% however, in author_list_en_cn.tex, there is a traditional chinese character for the name
%% S.~L.~Olsen(馬鵬)
%% so in there, we use the {\CJKfamily{bsmi}馬<E9> <B5><AC>} to show that name
%% and for the name 郭玥，the‘玥’can not be shown in gbsn,
%% and for the name 康晓珅， {\CJKfamily{bsmi}珅}
%% so we also use {\CJKfamily{bsmi}玥} to show it
%% If editors want to change the font, please notice that problem.

\fancyfoot[C]{\small \thepage}

\title{Measurement of $e^+e^- \rightarrow \DDbar$ Cross Sections at the $\psipp$ Resonance\thanks{
Supported in part by
National Key Basic Research Program of China under Contract No. 2015CB856700; National Natural Science Foundation of China (NSFC) under Contracts Nos. 11235011, 11335008, 11425524, 11625523, 11635010; the Chinese Academy of Sciences (CAS) Large-Scale Scientific Facility Program; the CAS Center for Excellence in Particle Physics (CCEPP); Joint Large-Scale Scientific Facility Funds of the NSFC and CAS under Contracts Nos. U1332201, U1532257, U1532258; CAS Key Research Program of Frontier Sciences under Contracts Nos. QYZDJ-SSW-SLH003, QYZDJ-SSW-SLH040; 100 Talents Program of CAS; National 1000 Talents Program of China; INPAC and Shanghai Key Laboratory for Particle Physics and Cosmology; German Research Foundation DFG under Contracts Nos. Collaborative Research Center CRC 1044, FOR 2359; Istituto Nazionale di Fisica Nucleare, Italy; Koninklijke Nederlandse Akademie van Wetenschappen (KNAW) under Contract No. 530-4CDP03; Ministry of Development of Turkey under Contract No. DPT2006K-120470; National Science and Technology fund; The Swedish Research Council; U. S. Department of Energy under Contracts Nos. DE-FG02-05ER41374, DE-SC-0010118, DE-SC-0010504, DE-SC-0012069; University of Groningen (RuG) and the Helmholtzzentrum fuer Schwerionenforschung GmbH (GSI), Darmstadt; WCU Program of National Research Foundation of Korea under Contract No. R32-2008-000-10155-0.}}

\maketitle

%-------- INSERT HERE ------------
\begin{small}
\begin{center}
M.~Ablikim(麦迪娜)$^{1}$, M.~N.~Achasov$^{9,d}$, S. ~Ahmed$^{14}$, M.~Albrecht$^{4}$, M.~Alekseev$^{55A,55C}$, A.~Amoroso$^{55A,55C}$, F.~F.~An(安芬芬)$^{1}$, Q.~An(安琪)$^{42,52}$, Y.~Bai(白羽)$^{41}$, O.~Bakina$^{26}$
% ...
\\
\vspace{0.2cm}
(BESIII Collaboration)\\
\vspace{0.2cm} {\it
$^{1}$ Institute of High Energy Physics, Beijing 100049, People's Republic of China\\
% ...
$^{4}$ Bochum Ruhr-University, D-44780 Bochum, Germany\\
% ...
$^{9}$ G.I. Budker Institute of Nuclear Physics SB RAS (BINP), Novosibirsk 630090, Russia\\
% ...
$^{14}$ Helmholtz Institute Mainz, Johann-Joachim-Becher-Weg 45, D-55099 Mainz, Germany\\
% ...
$^{26}$ Joint Institute for Nuclear Research, 141980 Dubna, Moscow region, Russia\\
% ...
$^{41}$ Southeast University, Nanjing 211100, People's Republic of China\\
$^{42}$ State Key Laboratory of Particle Detection and Electronics, Beijing 100049, Hefei 230026, People's Republic of China\\
% ...
$^{52}$ University of Science and Technology of China, Hefei 230026, People's Republic of China\\
% ...
$^{55}$ (A)University of Turin, I-10125, Turin, Italy; (B)University of Eastern Piedmont, I-15121, Alessandria, Italy; (C)INFN, I-10125, Turin, Italy\\
% ...
$^{d}$ Also at the Novosibirsk State University, Novosibirsk, 630090, Russia\\
% ...
}\end{center}

\vspace{0.4cm}
\end{small}
%-------- END INSERT ------------

\begin{abstract}
We report new measurements of the cross sections for the production of $D\bar{D}$ final states 
at the $\psipp$ resonance.  Our data sample consists of an integrated luminosity of 
2.93~fb$^{-1}$ of $e^+e^-$ annihilation data produced by the BEPCII collider
and collected and analyzed with the BESIII detector.  We exclusively reconstruct three $D^0$ and six $D^+$ 
hadronic decay modes and use the ratio of the yield of fully reconstructed $D\bar{D}$ events 
(``double tags") to the yield of all reconstructed $D$ or $\bar{D}$ mesons (``single tags")
to determine the number of $D^0\bar{D}^0$ and $D^+D^-$ events, benefiting 
from the cancellation of many systematic uncertainties.  Combining these yields with an independent 
determination of the integrated luminosity of the data sample, we find the cross sections to be 
$\sigma(e^+e^- \rightarrow D^0\bar{D}^0)=(3.615 \pm 0.010  \pm 0.038)$~nb and 
$\sigma(e^+e^- \rightarrow D^+D^-)=(2.830 \pm 0.011  \pm 0.026)$~nb,
where the uncertainties are statistical and systematic, respectively.
\end{abstract}



\footnotetext[0]{\hspace*{-3mm}\raisebox{0.3ex}{$\scriptstyle\copyright$}2013
Chinese Physical Society and the Institute of High Energy Physics
of the Chinese Academy of Sciences and the Institute
of Modern Physics of the Chinese Academy of Sciences and IOP Publishing Ltd}%

\begin{multicols*}{2}

\section{INTRODUCTION}

The $\psipp$ resonance is the lowest-energy charmonium state above the threshold for decay to charmed 
meson pairs.  The expectation that the $\psipp$ should decay predominantly to $D^0\bar{D}^0$ and 
$D^+D^-$ has been validated by experiment~\cite{Olive:2016xmw}, although inconsistent
results for the branching fraction of $\psipp$ to non-$D\bar{D}$ final states have been 
reported~\cite{Ablikim:2008zzb,Besson:2005hm}.  The cross 
sections $\sigma(e^+e^- \rightarrow D^0\bar{D}^0$) and $\sigma(e^+e^- \rightarrow D^+D^-$) at 
center-of-mass energy
$E_{\rm{cm}}=$3.773~GeV, the peak of the $\psipp$ resonance, can be measured precisely and 
are necessary input for normalizing some measurements of charmed meson properties in
$\psipp$ decays.  The most precise determinations to date are from the CLEO-c 
Collaboration~\cite{Bonvicini:2014ab} using 818~pb$^{-1}$ of $e^+e^-$ annihilation data at 
$E_{\rm{cm}}=3774 \pm 1$~MeV,
$\sigma(e^+e^- \rightarrow D^0\bar{D}^0)=(3.607 \pm 0.017  \pm 0.056)$~nb and
$\sigma(e^+e^- \rightarrow D^+D^-)=(2.882 \pm 0.018  \pm 0.042)$~nb.
In this paper we report measurements of the $D\bar{D}$ cross 
sections using fully reconstructed $D^0$ and $D^+$ mesons in a $\psipp$ data sample that is 
approximately $3.6$ times larger than CLEO-c's. Here and throughout this paper, charge-conjugate 
modes are implied unless explicitly stated.

Our procedure is an application of the $D$-tagging technique developed by the MARK~III 
Collaboration~\cite{Baltrusaitis:1985iw}, exploiting the kinematics of $D\bar{D}$ production 
just above threshold at the $\psipp$ resonance.  We use ratios of fully reconstructed $D$ 
mesons (``single tags'') and $D \bar{D}$ events (``double tags'') to determine the total numbers 
of $D \bar{D}$ pairs.  This procedure benefits from the cancellation of systematic uncertainties associated with 
efficiencies and input branching fractions, giving better overall precision than measurements based 
on single tags.  The production of $D^0\bar{D}^0$ pairs in a pure $C=-1$ state complicates the 
interpretation of measurements at $\psipp$ by introducing correlations between the $D^0$ and 
$\bar{D}^0$ decays.  We apply corrections derived by Asner and Sun~\cite{Asner:2005wf} to 
remove the bias introduced by these correlations.

\section{BESIII DETECTOR}

Our measurement has been made with the BESIII detector at the BEPCII collider of the Institute for 
High Energy Physics in Beijing.  Data were collected at the $\psipp$ peak, with
$E_{\rm{cm}}=3.773$~GeV.  The integrated luminosity of this sample has previously 
been determined with large-angle Bhabha scattering events to be 
2.93~fb$^{-1}$~\cite{Ablikim:2014gna,Ablikim:2015orh}, with an uncertainty of 0.5\% dominated 
by systematic effects.  An additional data sample of 44.9~pb$^{-1}$ at $E_{\rm{cm}}=3.650$~GeV has been 
used to assess potential background from continuum production under the $\psipp$.  

BESIII is a general-purpose magnetic spectrometer with a geometrical acceptance 
of 93\% of $4\pi$.  Charged particles are reconstructed in a 43-layer helium-gas-based drift chamber~(MDC), 
which has an average single-wire resolution of 135~$\mu$m.  A uniform axial magnetic field of 1~T is provided 
by a superconducting solenoid, allowing the precise measurement of charged particle trajectories.  The resolution 
varies as a function of momentum, and is 0.5\% at 1.0~GeV/$c$.   The MDC is also instrumented to measure 
the specific ionization ($dE/dx$) of charged particles for
particle identification.  Additional particle identification is provided by a time-of-flight system (TOF) constructed 
as a cylindrical (``barrel") structure with two 5-cm-thick plastic-scintillator layers and two ``end caps'' with one 
5-cm layer.  The time resolution in the barrel is approximately 80~ps, and in the end 
caps it is 110~ps.  Just beyond the TOF  is an electromagnetic calorimeter~(EMC) consisting of 6240 CsI(Tl) 
crystals, also configured as a barrel and two end caps.  For 1.0-GeV photons, the energy resolution is 2.5\% in 
the barrel and it is 5\% in the end caps.   This entire inner detector resides in the solenoidal magnet, which is 
supported by an octagonal flux-return yoke instrumented with resistive-plate counters interleaved with steel for 
muon identification (MUC).  More detailed information on the design and performance of the BESIII detector can 
be found in Ref.~\cite{Ablikim:2009aa}. 

\section{TECHNIQUE}

To select a $D\bar{D}$ event, we fully reconstruct a $D$ using tag modes that have sizable branching 
fractions and can be reconstructed with good efficiency and reasonable background.  We use three $D^0$ 
and six $D^+$ tag modes: $D^0 \to K^- \pi^+$, $D^0 \to K^- \pi^+ \pi^0$,  $D^0 \to K^- \pi^+ \pi^+ \pi^-$, 
$D^+ \to K^-\pi^+ \pi^+$,  $D^+ \to K^- \pi^+ \pi^+ \pi^0$, $D^+ \to K_S^0 \pi^+$, $D^+ \to K_S^0 \pi^+ \pi^0$, 
$D^+ \to K_S^0 \pi^+ \pi^+ \pi^-$, and $D^+ \to K^- K^+ \pi^+$. 

When both the $D$ and $\bar{D}$ in an event decay to tag modes we can fully reconstruct 
the entire event.  These double-tag events are selected when the event has two single tags and 
satisfies the additional requirements that the reconstructed single tags have opposite net charge, opposite-charm 
$D$ parents and no shared tracks.  The yield $X_{i}$ for single-tag mode $i$ is given by Eq.~\autoref{eq:stagnd}:

\begin{equation}
X_{i} = N_{D\bar{D}} \cdot \mathcal{B}(D\to i)\cdot \epsilon_{i},
\label{eq:stagnd}
\end{equation}

\noindent where $N_{D\bar{D}}$ is the total number of $D\bar{D}$ events,
$\mathcal{B}(D\to i)$ is the branching
fraction for decay mode $i$, and $\epsilon_{i}$ is the reconstruction
efficiency for the mode,  determined with Monte Carlo (MC) simulation.  Extending this reasoning, the 
yields for $\bar{D}$ decaying to mode $j$ and for $ij$ double-tag events, in which the $D$ decays to 
mode $i$ and the $\bar{D}$ decays to mode $j$, are given as follows:

\begin{equation}
Y_{j} = N_{D\bar{D}}\cdot \mathcal{B}(\bar{D}\to j)\cdot \epsilon_{j}
\label{eq:stagndbar}
\end{equation}
\noindent and
\begin{equation}
Z_{ij}=N_{D\bar{D}}\cdot \mathcal{B}(D\to
i)\cdot \mathcal{B}(\bar{D}\to j)\cdot \epsilon_{ij}.
\label{eq:dtagnd}
\end{equation}

\noindent In these equations, $Z_{ij}$ is the yield for the double-tag mode $ij$, and $\epsilon_{ij}$
is the efficiency for reconstructing both tags in the same event.  
Combining Eqs.~\autoref{eq:stagnd}, ~\autoref{eq:stagndbar} and ~\autoref{eq:dtagnd}, $N_{D\bar{D}}$ can 
be expressed as

\begin{equation}
N_{D\bar{D}} =
\frac{X_{i}\cdot Y_{j}\cdot \epsilon_{ij}}{Z_{ij}\cdot \epsilon_{i}\cdot \epsilon_{j}}.
\end{equation}

\noindent The cancellation of systematic uncertainties occurs through the ratio of efficiencies 
$\epsilon_{ij}/(\epsilon_i \cdot \epsilon_j)$. 
The measured $N_{D\bar{D}}$ from each combinations of $i$ and $j$ are then averaged, weighted by
their statistical uncertainties.
Finally, to determine cross sections we 
divide $N_{D\bar{D}}$ by the integrated 
luminosity $\mathcal{L}$ of the $\psipp$ sample, 
$\sigma(e^+e^- \rightarrow D\bar{D}) =N_{D\bar{D}}/{\mathcal{L}}$.

\section{PARTICLE RECONSTRUCTION}

Detection efficiencies and backgrounds for this analysis have been studied with detailed simulations of the BESIII 
detector based on GEANT4~\cite{Agostinelli:2002hh}.  High-statistics MC samples were produced for generic 
$D^0\bar{D}^0$ and $D^+D^-$ decays from $\psipp$, $\qqbar \to\text{light hadrons}$ $(q = u, d$ or $s)$, 
$\tau^+\tau^-$, and radiative return to $\jpsi$ and $\psip$.  
The $D^0\bar{D}^0$, $D^+D^-$, $\qqbar$, and $\tau^+\tau^-$ states were generated 
using KKMC~\cite{Jadach:1999vf,Jadach:2000ir}, while the $\gamma \jpsi$ and $\gamma \psip$ were generated 
with EvtGen~\cite{Lange:2001uf}.
All were then decayed with EvtGen, except for the $\qqbar$ and $\tau^+\tau^-$, which were modeled with
the LUNDCHARM~\cite{Chen:2000tv}
and the TAUOLA~\cite{taula26,Jadach:1999vf} generators, respectively.

Data and MC samples are treated identically for the selection of $D$ tags.  All particles used to 
reconstruct a candidate must pass requirements specific to the particle type.   Charged particles are required
to be within the fiducial region for reliable tracking ($|\cos\theta| < 0.93$, where $\theta$ is the polar angle
relative to the beam direction) and to pass within 1~cm (10~cm) of the interaction point in the plane transverse 
to the beam direction (along the beam direction).  Particle identification 
is based on TOF and $dE/dx$ measurements, with the identity as a pion or kaon assigned based on which 
hypothesis has the higher probability.  To be selected as a photon, an EMC shower must 
not be associated with any charged track~\cite{Ablikim:2015ycp},
must have an EMC hit time between $0$ and $700$~ns
to suppress activity that is not consistent with originating from the collision event,
must
have an energy of
at least $25$~MeV if it is in the barrel region of the detector ($|\cos\theta| < 0.8$), 
and $50$~MeV if it is in the end cap region ($0.84 < |\cos\theta| < 0.92$)
to suppress noise in the EMC as a potential background to real photons.
Showers in the transition region between the barrel and end cap are excluded.

$K_S^0$ mesons are reconstructed from the decay into $\pi^+\pi^-$.  Because of the cleanliness of the 
selection and the possibility of a measurably displaced decay vertex, the pions are not required
to pass the usual particle identification or interaction-point requirements.  A fit is performed
with the pions constrained to a common vertex and the $K_S^0$ candidate is accepted if the fit satisfies $\chi^2<100$ 
and the candidate mass is within $\sim3\sigma$ of the nominal $K_S^0$ mass 
($487-511$~MeV/$c^2$).  The momentum of the $K_S^0$ that is obtained from the constrained-vertex fit is used 
for the subsequent reconstruction of $D$-tag candidates.  $\pi^0$ mesons are reconstructed through the
decay into two photons.  Both photons for a $\pi^0$ candidate must pass the above selection criteria, and at 
least one of them must be in the barrel region of the detector.  To be accepted a $\pi^0$ candidate must 
have an invariant mass between $115$~MeV/$c^2$ and $150$~MeV/$c^2$.  The photons are then refitted with a $\pi^0$ 
mass constraint and the resulting $\pi^0$ momentum is used for the reconstruction of $D$-tag candidates.

\section{EVENT SELECTION}

In addition to the requirements on the final-state particles, the reconstructed $D$-tag candidates must pass 
several additional requirements that ensure the measured candidate energy and momentum are close to the expected 
values for production via $\psipp \to D\bar{D}$.  The first of these requirements is 
$\Delta E = E_D - E_{\rm{beam}}\simeq 0$, where $E_D$ is the energy of the reconstructed $D$ candidate and 
$E_{\rm{beam}}$ is the beam energy. 
In calculating $\Delta E$ we use the beam energy calibrated 
with $D^0$ and $D^+$ decays,
combining groups of nearby runs to obtain sufficient statistics.
 Selection requirements on $\Delta E$ are determined separately for each 
tag mode for data and MC to account for differing resolutions.  
As shown in Table~\ref{table:delecuts},
for modes decaying into all charged tracks, 
the requirements are set to $\pm 3\sigma$ about the mean, while for modes with a $\pi^0$, the requirements are asymmetric 
about the mean, extending on the low side to $-4\sigma$ to accommodate the tail from the photon energy resolution.
\begin{table*}[htbp]
\caption{The selected range on $\Delta E$ is $\pm3\sigma$ about the mean, except
  that for modes with a $\pi^0$ an extended lower bound of $-4\sigma$ is
  used. The resolutions and means are 
extracted by fitting with a double Gaussian, weighted by
the two Gaussian yields,
and
determined separately for data and MC. }
\label{table:delecuts}
\begin{center}
\small
\begin{tabular}{l|c|c|c|c}
\hline
\multicolumn{1}{c|}{} & \multicolumn{ 2}{c|}{MC} & \multicolumn{2}{c}{Data} \\ 
\hline
\multicolumn{1}{c|}{Tag mode} & \multicolumn{1}{c|}{$\sigma$ (MeV)} &
 \multicolumn{1}{c|}{Mean (MeV)} &
 \multicolumn{1}{c|}{$\sigma$ (MeV)} &
 \multicolumn{1}{c}{Mean (MeV)} \\ \hline
$D^0 \to K^- \pi^+$ & $7.6$ & $-0.4$ & $9.4$ & $-0.8$ \\ \hline
$D^0 \to K^- \pi^+ \pi^0$ & $14.1$ & $-7.6$ & $15.4$ & $-7.6$ \\ \hline
$D^0 \to K^- \pi^+ \pi^+ \pi^-$ & $8.2$ & $-1.4$ & $9.8$ & $-2.0$ \\ \hline
$D^+ \to K^- \pi^+ \pi^+$ & $7.2$ & $-0.9$ & $8.6$ & $-1.2$ \\ \hline
$D^+ \to K^- \pi^+ \pi^+ \pi^0$ & $12.8$ & $-6.9$ & $13.7$ & $-6.9$ \\ \hline
$D^+ \to K^{0}_{S} \pi^+$ & $6.7$ & $0.4$ & $8.4$ & $-0.1$ \\ \hline
$D^+ \to K^{0}_{S} \pi^+ \pi^0$ & $14.6$ & $-7.7$ & $16.2$ & $-7.9$ \\ \hline
$D^+ \to K^{0}_{S} \pi^+ \pi^+ \pi^-$ & $8.2$ & $-1.1$ & $10.4$ & $-1.7$ \\
 \hline
$D^+ \to K^+ K^- \pi^+$ & $6.2$ & $-1.1$ & $7.2$ & $-1.5$ \\
\hline
\end{tabular}
\end{center}
\end{table*}
Figure~\ref{fig:deltae} shows the data and MC overlays of the $\Delta E$ distributions by mode. 
\begin{center}
  \includegraphics[width=8.5cm]{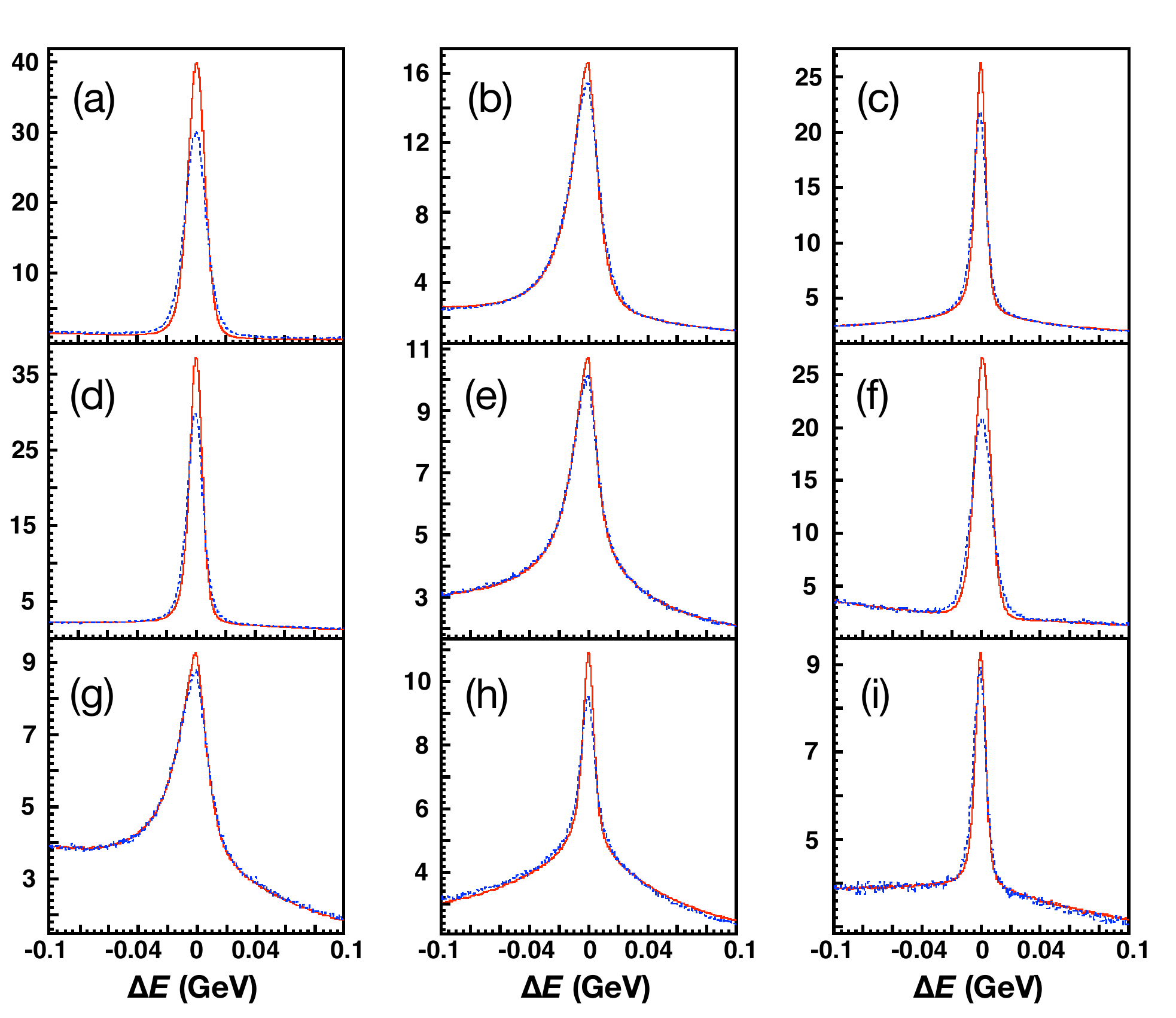}
\figcaption
{\label{fig:deltae}(color online) $\Delta E$ line shape for various single-tag mode (arbitrarily scaled). Starting from the top left, the modes are: (a) $D^0
  \to K^- \pi^+$, (b) $D^0 \to K^- \pi^+ \pi^0$, (c) $D^0
  \to K^- \pi^+ \pi^+ \pi^-$, (d) $D^+
  \to K^- \pi^+ \pi^+$, (e) $D^+ \to K^- \pi^+ \pi^+ \pi^0$,
  (f) $D^+ \to K_S^0 \pi^+$, (g) $D^+ \to K_S^0 \pi^+ \pi^0$,
  (h) $D^+ \to K_S^0 \pi^+ \pi^+ \pi^-$, and (i) $D^+ \to K^+
  K^- \pi^+$. These plots overlay the 3.773~GeV data (blue dashed histograms) and the corresponding
  narrower-width MC (red solid histograms).  Only requirements
  on the constituent particles and a very loose $\mbc$ requirement 
  ($1.83$~GeV/$c^2$ $\leq\mbc\leq 1.89$~GeV/$c^2$) have been applied.}
\end{center}

The second variable used in selecting $D$ tags is the beam-constrained mass 
$\mbc c^{2} = \sqrt{E_{\rm{beam}}^2 - |\textbf{p}_{\rm{tag}}c|^2}$, where $\textbf{p}_{\rm{tag}}$ is 
the 3-momentum of the candidate $D$.  We use $\mbc$  rather than the invariant mass 
because of the excellent precision with which the beam energy is known.  The requirement 
that $\mbc$ be close to the known $D$ mass ensures that the $D$ tag has the 
expected momentum.  After application of the $\Delta E$ requirement to single-tag candidates of 
a given mode, we construct an $\mbc$ distribution in the region of 
the known masses of charmed mesons
($1.83-1.89$~GeV/$c^2$).  For the MC a small upward shift of just under 1~MeV/$c$ is applied to the 
measured $D$ momentum for the calculation of $\mbc$ to compensate for input parameters 
that do not precisely match data.  Initial inspection of the distribution in data for the two-body mode 
$D^0  \to K^- \pi^+$ exhibited peaking near the high end of the $\mbc$ range not seen in MC.
We demonstrated this to be background from cosmic ray and QED events.  To eliminate it from the distribution,
additional requirements are applied in selecting $D^0  \to K^- \pi^+$ candidates with exactly two charged tracks.
We veto these events if they satisfy at least one of the following conditions: TOF information consistent 
with a cosmic ray event, particle identification information consistent with an $e^+e^-$ hypothesis, 
two tracks with EMC energy deposits consistent with an $e^+e^-$ hypothesis, or either track with particle identification and MUC information consistent with being a muon. 

\section{YIELDS AND EFFICIENCIES}

The $\mbc$ distribution for single-tag candidates for each mode is fitted 
with a MC-derived signal shape and an ARGUS function background~\cite{Albrecht:1990am}. 
The signal shape is convolved with a double Gaussian with a common mean to allow for differences 
in $\mbc$ resolution between data and MC.  Charge-conjugate modes are fitted
simultaneously with the double-Gaussian signal-shape parameters
constrained to be the same and the normalizations and background parameters allowed to
vary independently in the fit.  Peaking backgrounds contributed by decay modes that have similar final states
to the signal mode are included in the signal shape, although the yields are corrected after the fit to count
only true signal events.  

An example $\mbc$ fit is shown in Fig.~\ref{fig:fullparamstagfits}.
(The full set of fits is provided in the Appendix.)
In events with multiple single-tag candidates, the best candidate is chosen per mode and per charm to be
the one with the smallest $|\Delta E|$. 
Based on the fit results tight mode-dependent requirements on $\Delta E$ are applied.
To determine the tag yield, the $\mbc$ histogram is integrated within the signal region, 
$1.8580$ GeV/$c^2\le\mbc\le1.8740$ GeV/$c^2$ for $D^0$ modes and 
$1.8628$ GeV/$c^2\le\mbc\le1.8788$ GeV/$c^2$ for $D^+$ modes, and 
then the analytic integral of the ARGUS function in this region is subtracted.
The efficiency for each of the 18 single-tag modes is found by using MC truth information to 
determine the total number generated for the denominator and using the same cut-and-count 
method as used for data to determine the numerator. 
 The single-tag yields and efficiencies are 
summarized in Table~\ref{table:stagyieldseffs}, where
the efficiencies include branching fractions for $\pi^0\to\gamma\gamma$
and $K_S^0\to\pi^+\pi^-$ decays.
\begin{center}
  \includegraphics[width=8.5cm]{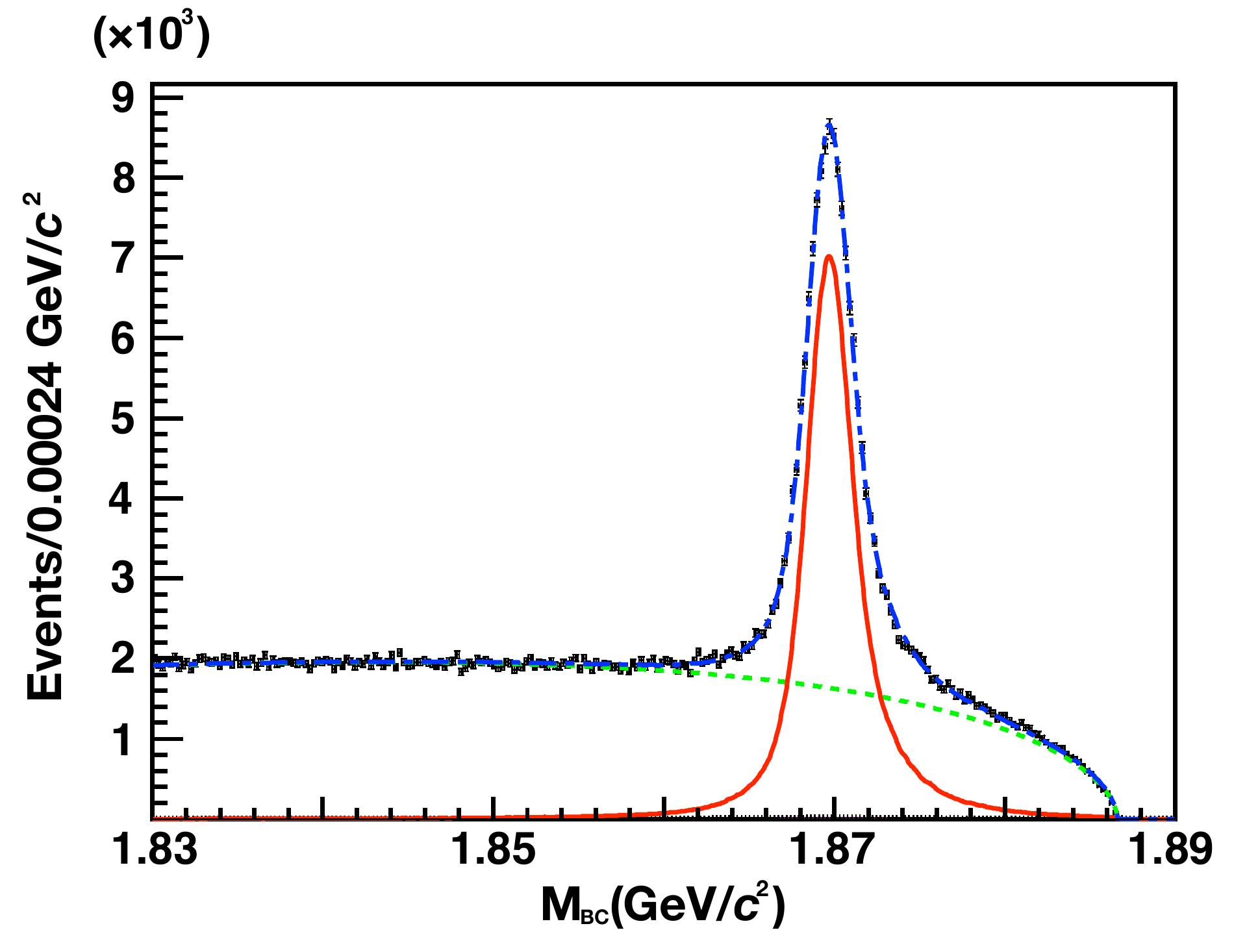}
\figcaption
{\label{fig:fullparamstagfits}(color online) $\mbc$ fit for single-tag mode $D^+ \to K^- \pi^+ \pi^+ \pi^0$, from data.
Blue dash-dot (green dashed) line represents the total fit (the fitted background shape)
and the red solid curve corresponds to the fitted signal shape.}
\end{center}

\begin{table*}[htbp]
\begin{center}
\tabcaption{\label{table:stagyieldseffs} Single-tag yields
after subtracting their corresponding peaking backgrounds
 from data and
  efficiencies from MC, as described in the text.
The uncertainties are statistical only.}
\scalebox{0.85}
{
\begin{tabular}{l|rcr|rcl||l|rcr|rcl}
\hline
\multicolumn{1}{c|}{Tag mode}             & \multicolumn{3}{c|}{Yield} & \multicolumn{3}{c||}{Efficiency (\%)} &
\multicolumn{1}{c|}{Tag mode}             & \multicolumn{3}{c|}{Yield} & \multicolumn{3}{c}{Efficiency (\%)} \\ \hline
$D^0 \to K^- \pi^+ $                            & $260,915$ & $\pm$ & $520$ & $63.125$ & $\pm$ &$0.007$ &
$\bar{D}^0 \to K^+ \pi^-$                    & $262,356$ & $\pm$ & $522$ & $64.272$ & $\pm$ & $0.006$ \\ \hline
$D^0 \to K^- \pi^+ \pi^0 $                   & $537,923$ & $\pm$ & $845$ & $35.253$ & $\pm$ & $0.007$ &
$\bar{D}^0 \to K^+ \pi^- \pi^0$           & $544,252$ & $\pm$ & $852$ & $35.761$ & $\pm$ & $0.007$ \\ \hline
$D^0 \to K^- \pi^+ \pi^+ \pi^- $         & $346,583$ & $\pm$ & $679$ & $38.321$ & $\pm$ & $0.007$ &
$\bar{D}^0 \to K^+ \pi^+ \pi^- \pi^-$ & $351,573$ & $\pm$ & $687$ & $39.082$ & $\pm$ & $0.007$ \\ \hline
$D^+ \to K^- \pi^+ \pi^+$                  & $391,786$ & $\pm$ & $653$ & $50.346$ & $\pm$ & $0.005$ &
$D^- \to K^+ \pi^- \pi^-$                   & $394,749$ & $\pm$ & $656$ & $51.316$ & $\pm$ & $0.005$ \\ \hline
$D^+ \to K^- \pi^+ \pi^+ \pi^0$         & $124,619$ & $\pm$ & $529$ & $26.138$ & $\pm$ & $0.014$ &
$D^- \to K^+ \pi^- \pi^- \pi^0$          & $128,203$ & $\pm$ & $539$ & $26.586$ & $\pm$ & $0.015$ \\ \hline
$D^+ \to K^{0}_{S} \pi^+$                      & $48,185$ & $\pm$ & $229$ & $36.726$ & $\pm$ & $0.008$ &
$D^- \to K^{0}_{S} \pi^-$                       & $47,952$ & $\pm$ & $228$ & $36.891$ & $\pm$ & $0.008$ \\ \hline
$D^+ \to K^{0}_{S} \pi^+ \pi^0$             & $114,919$ & $\pm$ & $471$ & $20.687$ & $\pm$ & $0.011$ &
$D^- \to K^{0}_{S} \pi^- \pi^0$              & $116,540$ & $\pm$ & $472$ & $20.690$ & $\pm$ & $0.011$ \\ \hline
$D^+ \to K^{0}_{S} \pi^+ \pi^+ \pi^-$   & $63,018$ & $\pm$ & $421$ & $21.966$ & $\pm$ & $0.019$ &
$D^- \to K^{0}_{S} \pi^+ \pi^- \pi^-$    & $62,982$ & $\pm$ & $421$ & $21.988$ & $\pm$ & $0.019$ \\ \hline
$D^+ \to K^+ K^- \pi^+$                     & $34,416$ & $\pm$ & $258$ & $41.525$ & $\pm$ & $0.042$ &
$D^- \to K^+ K^- \pi^-$                      & $34,434$ & $\pm$ & $257$ & $41.892$ & $\pm$ & $0.042$ \\ \hline
\end{tabular}
}
\end{center}
\end{table*}

Double tags are fully reconstructed events in which both the $D$ and the $\bar {D}$ pass the selection
criteria for one of the tag modes.  In events with multiple double-tag candidates, the best candidate per
mode combination per event is chosen 
with the $[\mbc(D) +  \mbc(\bar{D})]/2$ closest to the known $D$ mass.
Following a procedure similar to the single-tag counting, we fit the 
two-dimensional distribution of $\mbc(\bar{D})$ vs. $\mbc(D)$
for the selected single-tag modes to define the signal region for a cut-and-count determination of the 
double-tag yield.  A more sophisticated treatment of the background is required because of the correlations 
between the tags.  The signal shape is again derived from MC, using truth information and including peaking backgrounds with the signal.  
We found that convolving the MC shape with smearing functions to account for
the small data/MC resolution difference did not appreciably improve the accuracy of the tag yields, 
so no signal smearing is included in the double-tag fits.

The background shapes in the double-tag fits correspond to four possible ways of mis-reconstructing 
an event, as shown in Fig.~\ref{fig:twodsbsubtraction}. 
A direct product of a MC-derived signal shape with an analytic ARGUS function background, 
with shape parameters fixed to those of the corresponding single-tag fit, is used to represent
the background contributed by events with a correctly reconstructed $D$ and incorrectly reconstructed 
$\bar{D}$.  The background shape for the charm-conjugate case is similarly constructed.  For completely 
reconstructed continuum events or fully reconstructed but mispartitioned $D\bar{D}$ events (with particles 
assigned incorrectly to the $D$ and $\bar{D}$), a direct product of a double-Gaussian function and an 
ARGUS function rotated by $45^{\circ}$ is used. The kinematic limit and exponent parameters of the rotated
ARGUS function are fixed, while the slope parameter is allowed to be free in the fit.  Finally, the remaining background events 
with neither $D$ nor ${\bar D}$ correctly reconstructed are modeled with a direct product of two ARGUS
functions, with parameters taken from the corresponding single-tag fits.  An example fit to data is shown 
in Fig.~\ref{fig:detailedtagfit}. 
(The full set of fits is provided in the Appendix.)

\begin{center}
  \includegraphics[width=8.5cm]{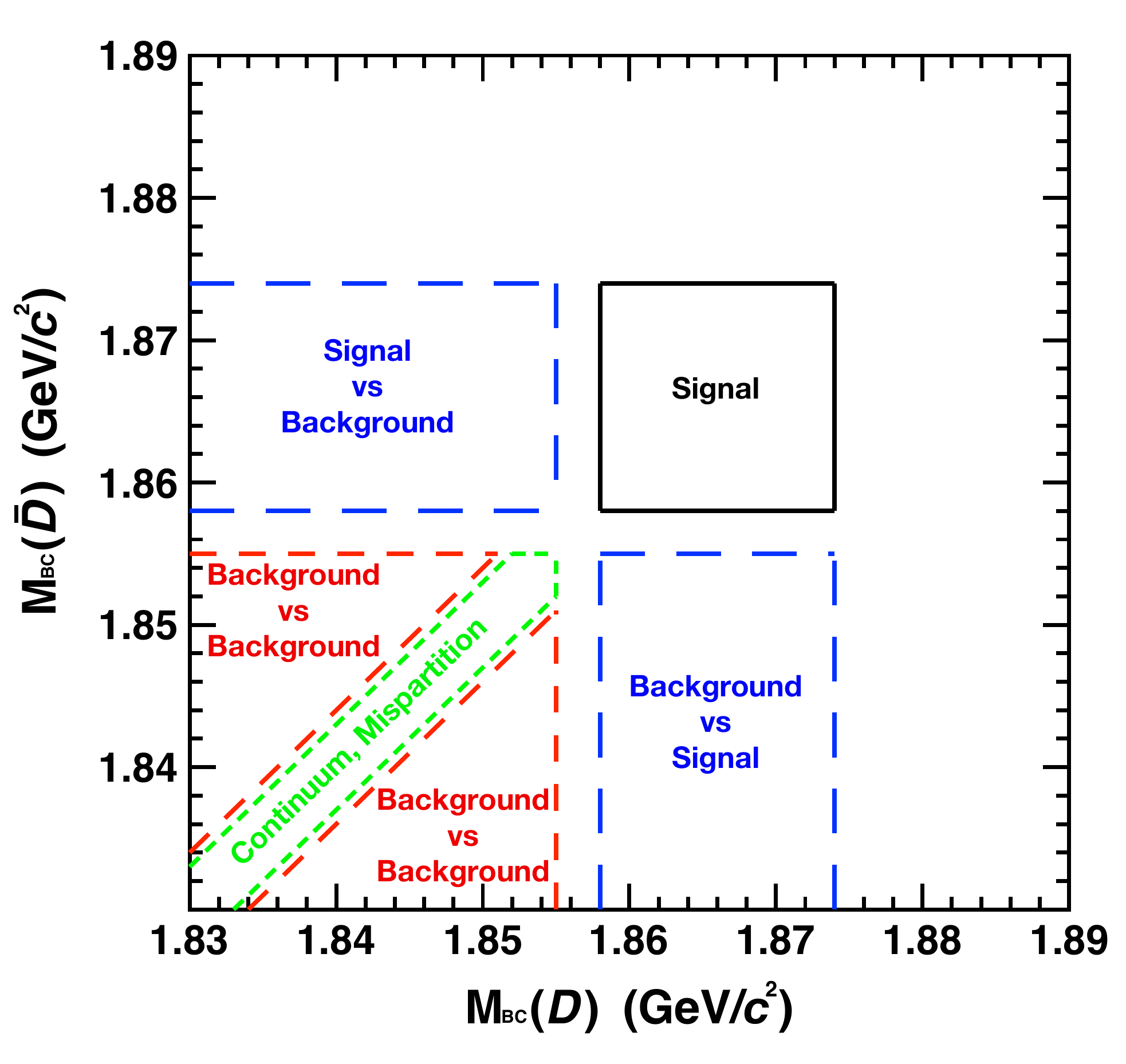}
\figcaption
{\label{fig:twodsbsubtraction}(color online)  The two-dimensional $\mbc$ plane divided into regions dominated by signal and
  various backgrounds. These regions represent the
  shapes used in the double-tag fitting method and sideband corrections described in the text.}
\end{center}

After the two-dimensional fit is performed, the $\mbc$
histogram is integrated within the same signal region as the single-tag
fits, and the integrals of the four background shapes are subtracted
from this total. 
The resultant double-tag yields and efficiencies, which
include branching fractions for $\pi^0\to\gamma\gamma$
and $K_S^0\to\pi^+\pi^-$ decays, are 
summarized in Tables~\ref{table:dtagd0yieldseffs} and \ref{table:dtagdpyieldseffs}.

\begin{table*}[htbp]
\caption{$D^0\bar{D}^0$ double-tag yields from data and
  efficiencies from MC, as described in the text.
The uncertainties are statistical only.}
\label{table:dtagd0yieldseffs}
\begin{center}
\begin{tabular}{l|rcr|rcl}
\hline
\multicolumn{1}{c|}{Tag mode} & \multicolumn{3}{c|}{Yield} & \multicolumn{3}{c}{Efficiency (\%)} \\ \hline
$D^0 \to K^- \pi^+ $~vs.~$ \bar{D}^0 \to K^+ \pi^-$ & $6,545$ & $\pm$ & $81$ &
$42.58$ & $\pm$&$0.13$ \\ \hline
$D^0 \to K^- \pi^+ $~vs.~$ \bar{D}^0 \to K^+ \pi^- \pi^0$ & $14,701$ & $\pm$ & $122$ &
$24.90$ & $\pm$ & $0.06$ \\ \hline
$D^0 \to K^- \pi^+ $~vs.~$ \bar{D}^0 \to K^+ \pi^+ \pi^- \pi^-$ &
$9,096$ & $\pm$ & $96$ & $25.54$ & $\pm$ & $0.08$ \\ \hline
$D^0 \to K^- \pi^+ \pi^0 $~vs.~$ \bar{D}^0 \to K^+ \pi^-$ & $14,526$ & $\pm$ & $122$ &
$24.94$ & $\pm$ & $0.06$ \\ \hline
$D^0 \to K^- \pi^+ \pi^0 $~vs.~$ \bar{D}^0 \to K^+ \pi^- \pi^0$ &
$30,311$ & $\pm$ & $176$ & $13.94$ & $\pm$ & $0.03$ \\ \hline
$D^0 \to K^- \pi^+ \pi^0 $~vs.~$ \bar{D}^0 \to K^+ \pi^+ \pi^- \pi^-$ &
$18,651$ & $\pm$ & $139$ & $14.35$ & $\pm$ & $0.03$ \\ \hline
$D^0 \to K^- \pi^+ \pi^+ \pi^- $~vs.~$ \bar{D}^0 \to K^+ \pi^-$ &
$8,988$ & $\pm$ & $96$ & $25.77$ & $\pm$ & $0.08$ \\ \hline
$D^0 \to K^- \pi^+ \pi^+ \pi^- $~vs.~$ \bar{D}^0 \to K^+ \pi^- \pi^0$ &
$18,635$ & $\pm$ & $139$ & $14.32$ & $\pm$ & $0.03$ \\ \hline
$D^0 \to K^- \pi^+ \pi^+ \pi^- $~vs.~$ \bar{D}^0 \to K^+ \pi^+ \pi^- \pi^-$
& $11,572$ & $\pm$ & $110$ & $14.86$ & $\pm$ & $0.04$ \\ \hline
\end{tabular}
\end{center}
\end{table*}

\begin{table*}[htbp]
\caption{$D^+D^-$ double-tag yields from data and efficiencies from
  MC, as described in the text.
The uncertainties are statistical only.}
\label{table:dtagdpyieldseffs}
\begin{center}
\begin{tabular}{l|rcr|rcl}
\hline
\multicolumn{1}{c|}{Tag mode} & \multicolumn{3}{c|}{Yield} & \multicolumn{3}{c}{Efficiency (\%)} \\ \hline
$D^+ \to K^- \pi^+ \pi^+ $~vs.~$ D^- \to K^+ \pi^- \pi^-$ & $18,800$ & $\pm$ & $138$ &
$26.02$ & $\pm$ & $0.05$ \\ \hline
$D^+ \to K^- \pi^+ \pi^+ $~vs.~$ D^- \to K^+ \pi^- \pi^- \pi^0$ &
$5,981$ & $\pm$ & $80$ & $13.62$ & $\pm$ & $0.05$ \\ \hline
$D^+ \to K^- \pi^+ \pi^+ $~vs.~$ D^- \to K^{0}_{S} \pi^-$ & $2,368$ & $\pm$ & $49$ &
$18.45$ & $\pm$ & $0.12$ \\ \hline
$D^+ \to K^- \pi^+ \pi^+ $~vs.~$ D^- \to K^{0}_{S} \pi^- \pi^0$ &
$5,592$ & $\pm$ & $75$ & $10.51$ & $\pm$ & $0.04$ \\ \hline
$D^+ \to K^- \pi^+ \pi^+ $~vs.~$ D^- \to K^{0}_{S} \pi^+ \pi^- \pi^-$ &
$2,826$ & $\pm$ & $53$ & $10.82$ & $\pm$ & $0.06$ \\ \hline
$D^+ \to K^- \pi^+ \pi^+ $~vs.~$ D^- \to K^+ K^- \pi^-$ & $1,597$ & $\pm$ & $40$ &
$20.87$ & $\pm$ & $0.15$ \\ \hline
$D^+ \to K^- \pi^+ \pi^+ \pi^0 $~vs.~$ D^- \to K^+ \pi^- \pi^-$ &
$6,067$ & $\pm$ & $80$ & $13.48$ & $\pm$ & $0.05$ \\ \hline
$D^+ \to K^- \pi^+ \pi^+ \pi^0 $~vs.~$ D^- \to K^+ \pi^- \pi^- \pi^0$ &
$1,895$ & $\pm$ & $53$ & $6.79$ & $\pm$ & $0.06$ \\ \hline
$D^+ \to K^- \pi^+ \pi^+ \pi^0 $~vs.~$ D^- \to K^{0}_{S} \pi^-$ & $693$ & $\pm$ & $26$
& $9.82$ & $\pm$ & $0.11$ \\ \hline
$D^+ \to K^- \pi^+ \pi^+ \pi^0 $~vs.~$ D^- \to K^{0}_{S} \pi^- \pi^0$ &
$1,726$ & $\pm$ & $44$ & $5.22$ & $\pm$ & $0.04$ \\ \hline
$D^+ \to K^- \pi^+ \pi^+ \pi^0 $~vs.~$ D^- \to K^{0}_{S} \pi^+ \pi^- \pi^-$
& $857$ & $\pm$ & $33$ & $5.41$ & $\pm$ & $0.06$ \\ \hline
$D^+ \to K^- \pi^+ \pi^+ \pi^0 $~vs.~$ D^- \to K^+ K^- \pi^-$ & $549$ & $\pm$ & $24$ &
$10.78$ & $\pm$ & $0.15$ \\ \hline
$D^+ \to K^{0}_{S} \pi^+ $~vs.~$ D^- \to K^+ \pi^- \pi^-$ & $2,352$ & $\pm$ & $48$ &
$18.96$ & $\pm$ & $0.12$ \\ \hline
$D^+ \to K^{0}_{S} \pi^+ $~vs.~$ D^- \to K^+ \pi^- \pi^- \pi^0$ & $722$ & $\pm$ & $27$
& $9.80$ & $\pm$ & $0.12$ \\ \hline
$D^+ \to K^{0}_{S} \pi^+ $~vs.~$ D^- \to K^{0}_{S} \pi^-$ & $269$ & $\pm$ & $16$ &
$13.95$ & $\pm$ & $0.27$ \\ \hline
$D^+ \to K^{0}_{S} \pi^+ $~vs.~$ D^- \to K^{0}_{S} \pi^- \pi^0$ & $678$ & $\pm$ & $26$
& $7.67$ & $\pm$ & $0.10$ \\ \hline
$D^+ \to K^{0}_{S} \pi^+ $~vs.~$ D^- \to K^{0}_{S} \pi^+ \pi^- \pi^-$ &
$383$ & $\pm$ & $20$ & $7.90$ & $\pm$ & $0.13$ \\ \hline
$D^+ \to K^{0}_{S} \pi^+ $~vs.~$ D^- \to K^+ K^- \pi^-$ & $191$ & $\pm$ & $14$ &
$15.2$ & $\pm$ & $0.34$ \\ \hline
$D^+ \to K^{0}_{S} \pi^+ \pi^0 $~vs.~$ D^- \to K^+ \pi^- \pi^-$ &
$5,627$ & $\pm$ & $75$ & $10.64$ & $\pm$ & $0.04$ \\ \hline
$D^+ \to K^{0}_{S} \pi^+ \pi^0 $~vs.~$ D^- \to K^+ \pi^- \pi^- \pi^0$ &
$1,708$ & $\pm$ & $43$ & $5.28$ & $\pm$ & $0.04$ \\ \hline
$D^+ \to K^{0}_{S} \pi^+ \pi^0 $~vs.~$ D^- \to K^{0}_{S} \pi^-$ & $624$ & $\pm$ & $25$
& $7.67$ & $\pm$ & $0.10$ \\ \hline
$D^+ \to K^{0}_{S} \pi^+ \pi^0 $~vs.~$ D^- \to K^{0}_{S} \pi^- \pi^0$ &
$1,557$ & $\pm$ & $40$ & $4.08$ & $\pm$ & $0.03$ \\ \hline
$D^+ \to K^{0}_{S} \pi^+ \pi^0 $~vs.~$ D^- \to K^{0}_{S} \pi^+ \pi^- \pi^-$
& $747$ & $\pm$ & $28$ & $4.26$ & $\pm$ & $0.05$ \\ \hline
$D^+ \to K^{0}_{S} \pi^+ \pi^0 $~vs.~$ D^- \to K^+ K^- \pi^-$ & $503$ & $\pm$ & $23$ &
$8.51$ & $\pm$ & $0.13$ \\ \hline
$D^+ \to K^{0}_{S} \pi^+ \pi^+ \pi^- $~vs.~$ D^- \to K^+ \pi^- \pi^-$ &
$2,857$ & $\pm$ & $53$ & $11.01$ & $\pm$ & $0.06$ \\ \hline
$D^+ \to K^{0}_{S} \pi^+ \pi^+ \pi^- $~vs.~$ D^- \to K^+ \pi^- \pi^- \pi^0$
& $924$ & $\pm$ & $34$ & $5.44$ & $\pm$ & $0.06$ \\ \hline
$D^+ \to K^{0}_{S} \pi^+ \pi^+ \pi^- $~vs.~$ D^- \to K^{0}_{S} \pi^-$ &
$313$ & $\pm$ & $18$ & $7.72$ & $\pm$ & $0.13$ \\ \hline
$D^+ \to K^{0}_{S} \pi^+ \pi^+ \pi^- $~vs.~$ D^- \to K^{0}_{S} \pi^- \pi^0$
& $778$ & $\pm$ & $29$ & $4.17$ & $\pm$ & $0.05$ \\ \hline
$D^+ \to K^{0}_{S} \pi^+ \pi^+ \pi^- $~vs.~$ D^- \to K^{0}_{S} \pi^+ \pi^-
\pi^-$ & $468$ & $\pm$ & $24$ & $4.28$ & $\pm$ & $0.06$ \\ \hline
$D^+ \to K^{0}_{S} \pi^+ \pi^+ \pi^- $~vs.~$ D^- \to K^+ K^- \pi^-$ &
$246$ & $\pm$ & $18$ & $8.96$ & $\pm$ & $0.19$ \\ \hline
$D^+ \to K^+ K^- \pi^+ $~vs.~$ D^- \to K^+ \pi^- \pi^-$ & $1,576$ & $\pm$ & $40$ &
$21.31$ & $\pm$ & $0.16$ \\ \hline
$D^+ \to K^+ K^- \pi^+ $~vs.~$ D^- \to K^+ \pi^- \pi^- \pi^0$ & $509$ & $\pm$ & $23$ &
$10.41$ & $\pm$ & $0.15$ \\ \hline
$D^+ \to K^+ K^- \pi^+ $~vs.~$ D^- \to K^{0}_{S} \pi^-$ & $185$ & $\pm$ & $14$ &
$14.48$ & $\pm$ & $0.33$ \\ \hline
$D^+ \to K^+ K^- \pi^+ $~vs.~$ D^- \to K^{0}_{S} \pi^- \pi^0$ & $468$ & $\pm$ & $22$ &
$8.23$ & $\pm$ & $0.13$ \\ \hline
$D^+ \to K^+ K^- \pi^+ $~vs.~$ D^- \to K^{0}_{S} \pi^+ \pi^- \pi^-$ &
$232$ & $\pm$ & $18$ & $8.62$ & $\pm$ & $0.19$ \\ \hline
$D^+ \to K^+ K^- \pi^+ $~vs.~$ D^- \to K^+ K^- \pi^-$ & $156$ &  $\pm$ &  $16$ &
$16.46$ & $\pm$ & $0.53$ \\ \hline
\end{tabular}
\end{center}
\end{table*}

We must correct the yields determined with the $\mbc$ fits 
(data and MC) for contributions from background processes that peak in the signal
region.  Such backgrounds come from other $D$ decays with similar kinematics and 
particle compositions as the specific signal mode.  We rely on MC, generated with 
world-average branching fractions~\cite{Olive:2016xmw}, to determine the fraction 
of peaking background events, as well as to calculate their selection efficiencies.  
We apply MC-determined corrections for these in every case where more than 0.01\% 
of the fitted yield is attributable to peaking background.   The largest contribution of 
peaking background is for $D^+ \to K^0_S \pi^+ \pi^+ \pi^-$, approximately 2.5\% 
of the fitted yield.  $D^0 \to K^- \pi^+ \pi^+ \pi^-$ and $D^+ \to K^0_S \pi^+ \pi^0$ both have 
$\sim 2.0\%$ of their fitted yields from peaking backgrounds, and all other modes have less
than $1.0\%$.  Because the peaking backgrounds come from well understood processes, like
doubly Cabibbo-suppressed modes, simultaneous misidentification of both a pion and
a kaon in an event, and charged pion pairs not from $K^0_S$ decays that pass the 
$K^0_S$ invariant mass requirement, we are confident that they are well modeled by the MC.  
 
The analysis described above results in a set of measured values of $N_{D\bar{D}{ij}}$, the
number of $D\bar{D}$ events determined with the single- and double-tag yields of positive 
tag mode $i$ and negative tag mode $j$.  The uncertainties are highly mode dependent 
because of branching fractions, efficiencies and backgrounds, so these measurements must be 
combined into an uncertainty-weighted mean taking into account correlations within and between the
mode-specific measurements.  We use an analytic procedure for this and demonstrated its 
reliability with a toy MC study.

\begin{center}
  \includegraphics[width=8.5cm]{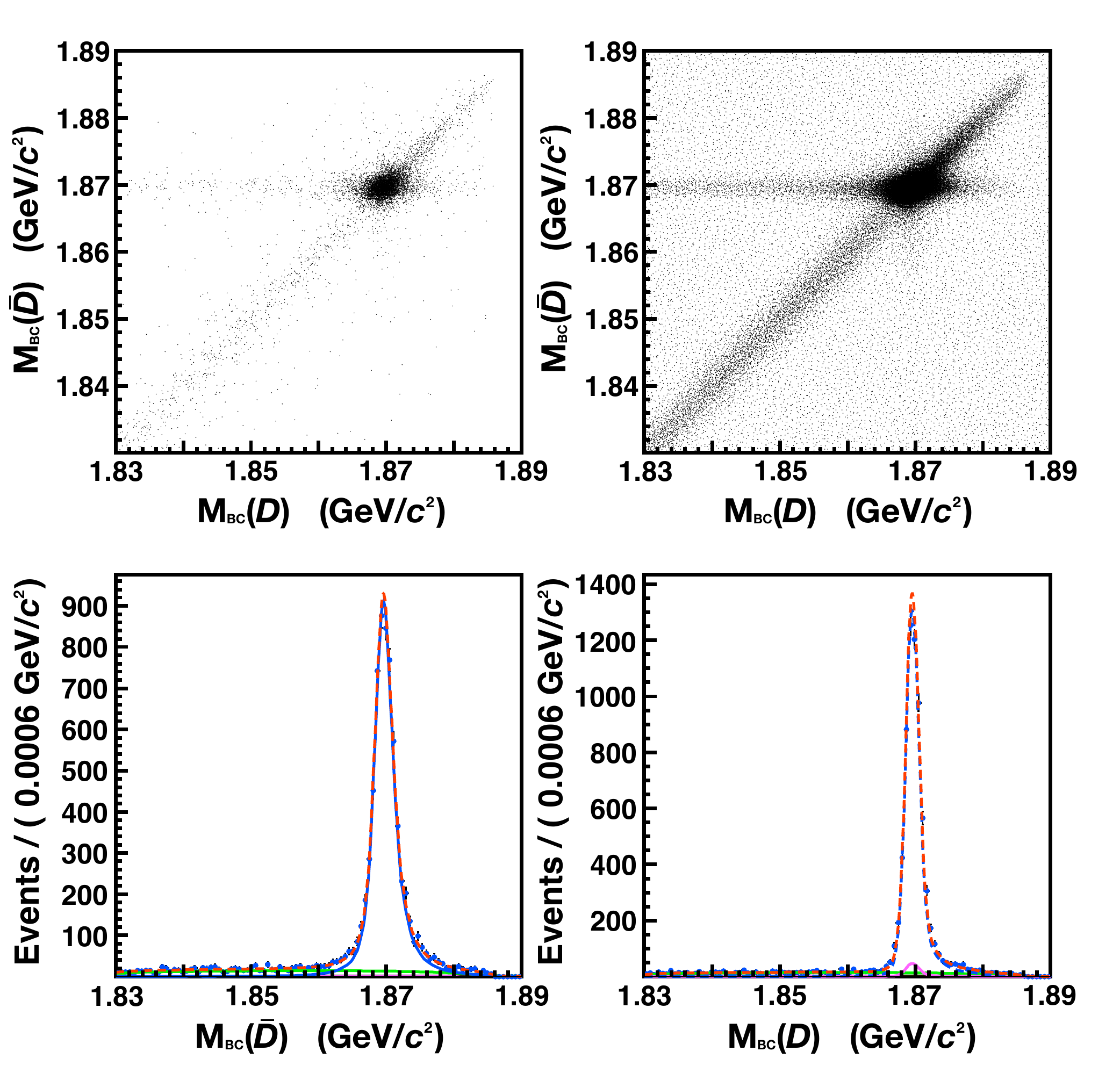}
\figcaption
{\label{fig:detailedtagfit}(color online)  Example two-dimensional $\mbc$ double-tag fit from data as described in
  the text, for tag mode  $K^+\pi^-\pi^-$ vs. $K^-\pi^+\pi^+\pi^0$.
  The top left figure is a scatter plot of the data and
  the top right is a scatter plot of the fit to the data.
  The bottom
  two plots are overlays of data and the fit projected onto the positive
  and negative charm $\mbc$ axes.
  The red dashed (blue solid) lines represent the total fits (the fitted signal shapes) and
  the solid green curves are the fitted background shapes.
  The magenta curve corresponds to the case when 
 $D^-\to K^+\pi^-\pi^-$ is reconstructed correctly, while
$D^+\to K^-\pi^+\pi^+\pi^0$ is not.}
\end{center}

For our full $2.93\mathrm{~fb}^{-1}$ $\psipp$ data sample we find
$N_{D^0\bar{D}^0}=(10,621\pm29)\times10^3$ and
$N_{D^+D^-}=(8,296\pm31)\times10^3$.  Using the integrated luminosity from
Ref.~\cite{Ablikim:2015orh}, we obtain observed cross sections for $D {\bar D}$ production
at the $\psipp$ of $\sigma(e^+e^- \rightarrow D^0\bar{D}^0)=(3.623 \pm 0.010)$~nb and 
$\sigma(e^+e^- \rightarrow D^+D^-)=(2.830 \pm 0.011)$~nb. 
Here, the uncertainties are statistical only.
The summed $\chi^2$ values relative to the mean for all pairs of 
tag modes are $13.2$ for $D^0\bar{D}^0$ ($9$ modes) and 
$53.6$ for $D^+D^-$ ($36$ modes).

We verified the reliability of our yield measurements with an ``In vs. Out'' test with MC by 
randomly partitioning our MC (signal and background) into ten 
statistically independent data-sized sets.  We  
determined single- and double-tag yields for these subsamples, calculated the $N_{D\bar{D}}$
and compared these to the true values for each.  The overall $\chi^2$ for these ten tests
was $10.7$ for $N_{D^0\bar{D}^0}$ and $12.4$ for $N_{D^+D^-}$, demonstrating 
that our procedure reliably determines both $N_{D\bar{D}}$ and its statistical uncertainty.  In a second 
test, the data sample was partitioned in time into five subsamples of approximately 0.5~fb$^{-1}$ 
each and measured $\sigma(e^+e^- \rightarrow D^0\bar{D}^0)$ and
$\sigma(e^+e^- \rightarrow D^+D^-)$ for each.   The values of $\chi^2$ for the hypothesis of equal values 
for all intervals were $5.4$ and $6.0$, respectively.

\section{EFFECTS OF QUANTUM CORRELATIONS}

As mentioned earlier in this paper, the $D^0 \bar {D}^0$ yield and cross section must be corrected for 
correlations introduced by production through a pure $C=-1$ state at the $\psipp$.  Asner and 
Sun~\cite{Asner:2005wf} provide correction factors that can be applied directly to our measured yields
with Eq.~\autoref{eq:qceqn1} for $D^0\to f$ and $\bar{D}^0\to f'$ and Eq.~\autoref{eq:qceqn2}
for the case $f=f'$.
\begin{equation}
N_{D^0\bar{D}^0}^{\rm{measured}} = N_{D^0\bar{D}^0}^{\rm{true}}\times
(1+r_f\tilde{y}_f+r_{f'}\tilde{y}_{f'}+r_fr_{f'}v^-_{ff'})\\
\label{eq:qceqn1}
\end{equation}
\begin{equation}
N_{D^0\bar{D}^0}^{\rm{measured}} = N_{D^0\bar{D}^0}^{\rm{true}}\times(1+2r_f\tilde{y}_f-r^2_f(2-z^2_f))\\
\label{eq:qceqn2}
\end{equation}
The quantities appearing in these equations can be expressed in terms of measured parameters of
$D^0$ decays and $D^0\bar{D}^0$ mixing, with 
$v^-_{jk} = (z_jz_k - w_jw_k)/2$, 
$z_j = 2\cos{\delta_j}$ and 
$w_j = 2\sin{\delta_j}$.  
$r_j$ and $\delta_j$ are defined by 
$\langle j|\bar{D}^0\rangle/\langle j|D^0\rangle = r_je^{i\delta_j}$, where
$r_j = |\langle j|\bar{D}^0\rangle/\langle j|D^0\rangle|$,
and $\delta_j$ is the average strong phase difference for the Cabibbo-favored tag mode.
The usual mixing parameters $x$ and $y$, which are related to the differences in masses
and lifetimes of the two mass eigenstates, enter through 
$\tilde{y}_j = y\cos{\delta_j} + x\sin{\delta_j}$.
The $D^0\to K^-\pi^+\pi^0$ and $K^-\pi^+\pi^+\pi^-$ tag modes require a slightly more complicated treatment because 
they are mixtures of modes with different phases.  This requires introducing coherence factors $R_j$ to 
characterize the variation of $\delta_j$ over phase space, with $z_j$ and $w_j$ being redefined as
$z_j = 2R_j\cos{\delta_j}$ and $w_j = 2R_j\sin{\delta_j}$~\cite{TQCA2}.

Table~\ref{tab:qcinputs} shows the input parameters that are used to obtain the correction factors 
and Fig.~\ref{fig:D0xsec_eachmodes} shows the corrections to $\sigma(e^+e^- \rightarrow D^0\bar{D}^0)$ for each of 
the nine double-tag modes, along with the average.  The overall effect is a relative change in 
$N_{D^0\bar{D}^0}$ of approximately $-0.2\%$, with final corrected values of 
$N_{D^0\bar{D}^0} = (10,597 \pm 28) \times 10^3$ and
$\sigma(e^+e^- \rightarrow D^0\bar{D}^0)=(3.615 \pm 0.010)$~nb. 
The uncertainties are statistical only.
The summed $\chi^2$ value relative to the mean for all pairs of 
tag modes is $11.8$ for $D^0\bar{D}^0$ ($9$ modes).

\begin{center}
\tabcaption{\label{tab:qcinputs} Input parameters for the quantum correlation corrections.
}
\def\1#1#2#3{\multicolumn{#1}{#2}{#3}}

\scalebox{1.0}
{
\begin{tabular}{l l  }
\hline
 $x = 0.0037\pm0.0016$         & \cite{hfag}\\
 $y = 0.0066^{+0.0007}_{-0.0010}$ & \cite{hfag}\\
 $r_{K\pi}^2 = 0.00349\pm0.00004$ & \cite{hfag}\\
 $\delta_{K\pi} = (11.8^{+9.5}_{-14.7})$\degree & \cite{hfag}\\
 $r_{K\pi\pi^0} = 0.0447\pm0.0012$ & \cite{UKCLEO} \\
 $\delta_{K\pi\pi^0} = (198^{+14}_{-15})$\degree (*) & \cite{UKCLEO} \\
 $R_{K\pi\pi^0} = 0.81\pm0.06$ & \cite{UKCLEO} \\
 $r_{K3\pi} = 0.0549\pm0.0006$ & \cite{UKCLEO} \\
 $\delta_{K3\pi} = (128^{+28}_{-17})$\degree (*) & \cite{UKCLEO} \\
 $R_{K3\pi} = 0.43^{+0.17}_{-0.13} $ & \cite{UKCLEO} \\
\hline
\1{2}{l}{(*) $180$\degree difference in} \\
\1{2}{l}{phase convention from Ref.~\cite{hfag}.}\\
\end{tabular}
}
\end{center}
\begin{center}
\includegraphics[keepaspectratio=true,width=8.5cm,angle=0]{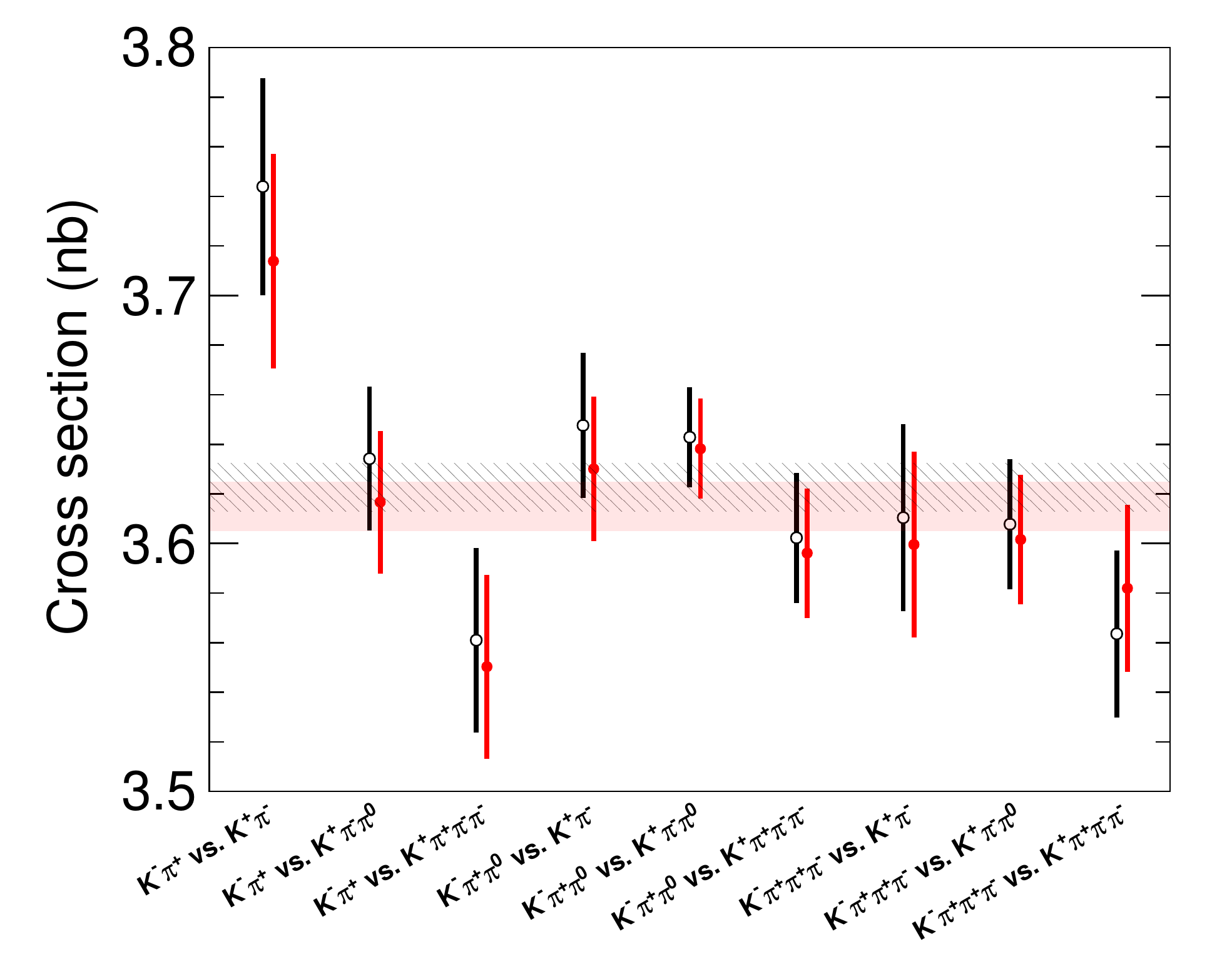}\\
\figcaption{\label{fig:D0xsec_eachmodes}(color online) $\sigma(e^+e^- \rightarrow D^0\bar{D}^0)$ 
for the nine double-tag modes, as labeled on the horizontal axis.  
The red (black) points show the $D^0\bar{D}^0$ cross section values with (without) the 
quantum correlation correction.  The light red (black shaded) band denotes the one-standard-deviation 
bound of the weighted average of the corrected  (uncorrected) measurements.}
\end{center}

\section{SYSTEMATIC UNCERTAINTIES}

The sources of systematic uncertainty that have been considered for the $D^0\bar{D}^0$ and $D^+D^-$
cross section measurements are listed in Table~\ref{table:systematics}.  

\begin{table*}[htbp]
\caption{Systematic uncertainties in the cross section measurements in \%.}
\label{table:systematics}
\begin{center}
\begin{tabular}{l|c|c}
\hline
  \multicolumn{1}{c|}{Source}  & $\sigma(e^+e^- \to D^0\bar{D}^0)$ & $\sigma(e^+e^- \to D^+D^-)$ \\
  \hline
Multiplicity-dependent efficiency & $0.4$ & $0.1$ \\
Other-side multiplicity & $<0.01$ & $0.22$ \\
Best-candidate selection & $0.45$ & $0.07$ \\
Single tag fit background shape          & $0.54$ & $0.64$ \\
Single tag fit signal shape                   & $0.26$ & $0.19$ \\
Double tag fit                                        & $0.28$ & $0.19$ \\
Cosmic/lepton veto                  & $0.06$ & N/A \\
$\psi(3770)$ line shape for ISR    & $0.15$ & $0.25$ \\
FSR simulation                         & $0.11$ & $0.10$ \\
Quantum correlation correction   & $0.2$ & N/A \\
Integrated luminosity                    & $0.5$ & $0.5$ \\
 \hline
Total   &   $1.05$ & $0.93$ \\
\hline              
\end{tabular}
\end{center}
\end{table*}

The double-tag technique 
used to determine the event yields and cross sections  $\sigma(e^+e^- \rightarrow D^0\bar{D}^0)$
and
$\sigma(e^+e^- \rightarrow D^+D^-)$
has the benefit of substantial cancellation of systematic uncertainties. 
Detector effects including tracking, particle identification, and $\pi^0$ and $K_S^0$ reconstruction, along 
with tag-mode resonant substructure and the $\Delta E$ requirement, all affect both single and double tags.  

There are, however, event-dependent effects that do not cancel 
in the 
efficiency ratio $\epsilon_{ij}/(\epsilon_i \cdot \epsilon_j)$. The event environment in which $D$ mesons are 
tagged affects the efficiency because higher multiplicities of charged tracks or $\pi^0$s lower the tagging efficiency. 
This can arise due to three possible sources: 
(1) differences in multiplicity-dependent efficiencies between data and MC,
(2) differences between the other-side multiplicities in data and MC due to imperfect 
knowledge of $D$ meson decay modes and rates, and
(3) sensitivity of the best candidate selection to the number of fake-tag background events.

To assess  a possible uncertainty due to the first source, 
we study efficiencies of tracking and particle identification for charged pions and kaons, as well as $\pi^0$ reconstruction,
based on doubly tagged $D^0\bar{D^0}$ and $D^+D^-$ samples. We estimate uncertainties  while observing 
how well our MC simulates these efficiencies in data
with different particle multiplicities.

We evaluate the effect of the second source for both tracks and $\pi^0$s by reweighting the MC to 
better match the multiplicities in data.
In this we assume that data and MC are consistent 
in the single track and $\pi^0$ reconstruction efficiencies.
We obtain corrected efficiencies separately for each tag mode, and
the difference with the nominal efficiency is used as
the systematic uncertainty.  The effect is larger for tag modes with greater multiplicity, and so the 
overall effect on $D^+D^-$ is greater than that on $D^0\bar{D}^0$.

The third source arises due to the fact that
we resolve multiple-candidate events when choosing single tags based on the smallest $|\Delta E|$.
This selection is imperfect and sometimes the wrong candidate is chosen, lowering the efficiency for
multiple-candidate events relative to single-candidate events.  Although a best-candidate selection 
is also applied to double tags, the number of multiple candidates in this case is small and the selection 
based on two beam-constrained masses is more reliable, so only the systematic uncertainty of 
best-candidate selection for single tags is considered.  Such uncertainty only arises when both the 
multiple-candidate rate is different between data and MC and the single- and multiple-candidate 
efficiencies are different.  These quantities can be measured both in data and MC, and the observed 
differences are propagated through to the systematic uncertainties in the cross sections.

Even though we fit both single and double tags to obtain the
yields and efficiencies, the differences between one- and two-dimensional fits and the much lower 
background levels of the double-tag $\mbc$ distributions limit the  cancellation.  We consider 
several variations of the fitting procedures and use the changes in efficiency-corrected yields to
estimate the systematic uncertainties. 

The uncertainty due to the single-tag background shape is probed by substituting a MC-derived 
background for the ARGUS function.  The uncertainty due to the signal shape is assessed by altering
the smearing of the MC-derived shape (single-Gaussian-convolved instead of the double-Gaussian-convolved). 
To assess the uncertainty in the double-tag fitting procedure, we obtain double-tag yields and 
efficiencies with an alternative sideband-subtraction method, dividing the two-dimensional $\mbc$
plane into sections representing the signal and various background
components, as shown in Fig.~\ref{fig:twodsbsubtraction}.  The signal area is the same as that used 
when fitting.  Horizontal and vertical bands are used to represent combinations with one correctly 
and one incorrectly reconstructed $D$; a diagonal band represents the background from completely 
reconstructed continuum events or mispartitioned $D\bar{D}$ events; and two triangles are used to 
represent the remaining background, which is mostly flat.  An estimate of the flat background is scaled by the 
ratios of the sizes of each of the other background regions and subtracted to obtain estimates of the 
non-flat backgrounds. These backgrounds are then scaled with area and ARGUS background parameters 
obtained from single-tag fits to determine the overall background subtraction and yield for the signal region 
for a specific tag mode.  The difference in efficiency-corrected double-tag yields for each mode between 
this method and the standard procedure is taken as the systematic uncertainty associated with the 
double-tag fitting method.

The cosmic and lepton veto suppresses cosmic ray and QED background in the single-tag selection
for the $D^0\to K^-\pi^+$ mode.  A cosmic ray background event is produced
by a single particle that is incorrectly reconstructed as two oppositely
charged tracks.  The net momentum of the two tracks is therefore close to zero, and 
typical QED events also have small net momentum. This small momentum produces 
$\mbc$ values close to the beam energy, so that residual cosmic ray and QED events 
passing the veto distort the $\mbc$ distribution.  Because the
processes responsible are not included in our MC samples or well described by the ARGUS 
background function, the fit results may be affected. 
To assess this effect, we performed alternative single-tag fits for $D^0\to K^-\pi^+$ with a 
cut-off in $\mbc$ at $1.88$~GeV/$c^2$, excluding the range where cosmic and QED 
events can contribute.  We found the resulting difference from the standard fit 
procedure to be $0.18\%$, which we take as the systematic uncertainty due to this effect.

The line shape of the $\psipp$ affects our analysis through the modeling of initial-state radiation 
(ISR) at the peak of the resonance.  The cross section for $\psipp$ production in radiative events 
depends on the cross section value at the lower effective $E_{\text{cm}}$ that results from 
ISR.   While this may partially cancel in the ratio, we treat it separately for single and double tags
because yields and efficiencies are affected with opposite signs, and because correlations are 
introduced for the double-tag fits that are not present in the single-tag fits.  The MC-determined 
efficiencies are affected through the $\Delta E$ requirements, which select against large ISR because  
the $\Delta E$ calculation assumes that the energy available to the $D$ is the full beam 
energy.  The data yields are affected via the $\mbc$ fit shape, which acquires an 
asymmetric high-side tail through the contribution of $\psipp$ production via ISR.  
More ISR causes a larger high-side tail in both the single- and double-tag signal shapes. 
Additionally, because both $D$ mesons lose energy when ISR occurs, double-tag 
events that include ISR will have a correlated shift in $\mbc$, causing such events to 
align with the diagonal to the high-side of the signal region in the two-dimensional 
$\mbc$ plane.  We use a preliminary BESIII measurement of the $\psipp$ line shape to 
re-weight the MC and repeat the $D$-counting procedure.  Combining the mode-by-mode 
variations in $N_{D\bar{D}}$ leads to the systematic uncertainty associated with the $\psipp$
line shape given in Table~\ref{table:systematics}.

The MC modeling of final-state radiation (FSR) may lead to a systematic difference 
between data and MC tag-reconstruction efficiencies.  FSR affects our measurement from 
the tag-side, so any systematic effect will also have some cancellation. To assess the uncertainty
due to FSR we created signal MC samples with and without modeling of FSR and measured the
changes in tag reconstruction efficiencies.  The largest difference was for $D^0 \to K^- \pi^+$, 
where the relative change in single-tag reconstruction efficiency was $~4\%$.
The $D^0 \to K^- \pi^+,
\bar{D}^0 \to K^+ \pi^-$ double-tag reconstruction efficiency also changed when FSR was turned
off, but the cancellation was not complete, with the ratio of efficiencies changing by  $1.2\%$.  Because
the variation of turning on and off FSR modeling is judged to be too extreme (FSR definitely happens), 
we take $25\%$ of this difference as our systematic uncertainty due to FSR modeling, a $0.3\%$ relative uncertainty on 
the MC reconstruction efficiency ratio.  To be conservative, we take the largest change, for the 
$D^0 \to K^- \pi^+$ mode, as the systematic uncertainty for all modes. 

The correction in the $D^0\bar{D}^0$ cross section due to the treatment of quantum correlations 
incurs systematic uncertainty associated with the parameters $x$, $y$, $\delta_{K\pi}$, and $r_{K\pi}^2$,
for which Ref.~\cite{hfag} provides correlation coefficients.  Ref.~\cite{UKCLEO} provides a similar coefficient 
table for the rest of the variables.  In evaluating our systematic uncertainty, 
we have doubled the reported uncertainties and treated them incoherently.  
Toy MC calculations were used to propagate these
uncertainties to $N_{D^0\bar{D}^0}$, giving a systematic uncertainty in the $D^0\bar{D}^0$ cross section
of $0.2\%$.

Finally, for the calculation of cross sections, the relative systematic uncertainty due to the integrated 
luminosity measurement is determined in Ref.~\cite{Ablikim:2014gna,Ablikim:2015orh} to be $0.5\%$.

\section{RESULTS AND CONCLUSIONS}

The separate sources of systematic uncertainty given in Table~\ref{table:systematics} are combined, 
taking correlations among them into account, to give overall systematic uncertainties in the 
$D^0\bar{D}^0$ and $D^+D^-$ cross sections of $1.05\%$ and $0.93\%$, respectively.  Including these 
systematic uncertainties, the final results of our analysis are as follows:
\begin{equation*}
\begin{array}{r c l}
N_{D^0\bar{D}^0}  & = & (10,597\pm28\pm98)\times10^3, \\
N_{D^+D^-}  & =  & (8,296\pm31\pm65)\times10^3, \\
\sigma(e^+e^-\to D^0\bar{D}^0) & = & (3.615\pm0.010\pm0.038)~\rm{nb}, \\
\sigma(e^+e^-\to D^+D^-) & = & (2.830\pm0.011\pm0.026)~\rm{nb}, \\
\sigma(e^+e^-\to D\bar{D}) & = & (6.445\pm0.015\pm0.048)~\rm{nb}, \\
\multicolumn{3}{c}{\text{and}}\\
\multicolumn{3}{l}{\sigma(e^+e^-\to D^+D^-)/\sigma(e^+e^-\to D^0\bar{D}^0)} \\
& = & (78.29\pm0.36\pm0.93)\%, \\
\end{array}
\end{equation*}
\noindent where the uncertainties are statistical and systematic, respectively.
In the determinations of $\sigma(e^+e^-\to D\bar{D})$
and $\sigma(e^+e^-\to D^+D^-)/\sigma(e^+e^-\to D^0\bar{D}^0)$,
the uncertainties of the charged and neutral cross sections
are mostly uncorrelated, except the systematic uncertainties due to
the assumed $\psi(3770)$ line shape, the FSR simulation, and the measurement
of the integrated luminosity.

In conclusion, we have used 2.93~fb$^{-1}$ of $e^+e^-$ annihilation data at the $\psipp$ resonance
collected by the BESIII detector at the BEPCII collider to measure the cross sections 
for the production of $D^0\bar{D}^0$ and $D^+D^-$.  The technique is full reconstruction of three 
$D^0$ and six $D^+$ hadronic decay modes and determination of the number of 
$D^0\bar{D}^0$ and $D^+D^-$ events using the ratio of single-tag and double-tag yields.  We find 
the cross sections to be $\sigma(e^+e^- \rightarrow D^0\bar{D}^0)=(3.615 \pm 0.010  \pm 0.038)$~nb and 
$\sigma(e^+e^- \rightarrow D^+D^-)=(2.830 \pm 0.011  \pm 0.026)$~nb,
where the uncertainties are statistical and systematic, respectively.  These results are consistent
with and more precise than the previous best measurement by the CLEO-c 
Collaboration~\cite{Bonvicini:2014ab} and
are necessary input for normalizing some measurements of charmed meson properties in $\psi(3770)$ decays.

\acknowledgments{The authors are grateful to Werner Sun of Cornell University for very helpful discussions. 
The BESIII collaboration thanks the staff of BEPCII and the computing center for their hard efforts.}

\end{multicols}

\section*{Appendix}

\begin{center}
	\includegraphics[width=12.5cm]{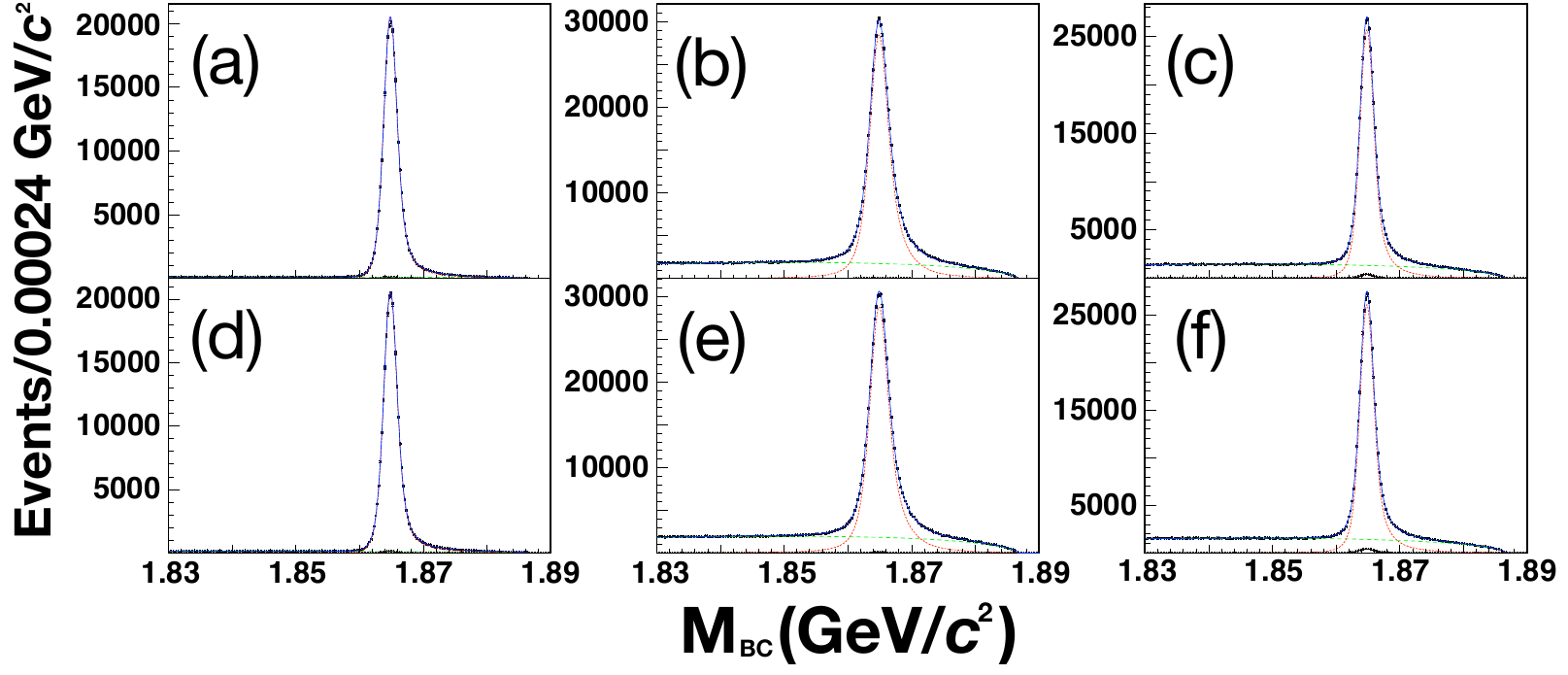}
	\figcaption{\label{std0} $\mbc$ fits for single-tag modes;
		(a) $D^0\to K^-\pi^+$, (b) $D^0\to K^-\pi^+\pi^0$, (c) $D^0\to K^-\pi^+\pi^+\pi^-$,
		(d) $\bar{D}^0\to K^+\pi^-$, (e) $\bar{D}^0\to K^+\pi^+\pi^-\pi^-$,
		(f) $\bar{D}^0\to K^+\pi^+\pi^-\pi^-$.
		Blue solid, red dotted, and green dashed  lines represent the total fits,
		the fitted signal shapes, and the fitted background shapes, respectively,
		while black histograms correspond to the expected peaking background components.}
\end{center}
\vspace{2mm}
\begin{center}
	\includegraphics[width=12.5cm]{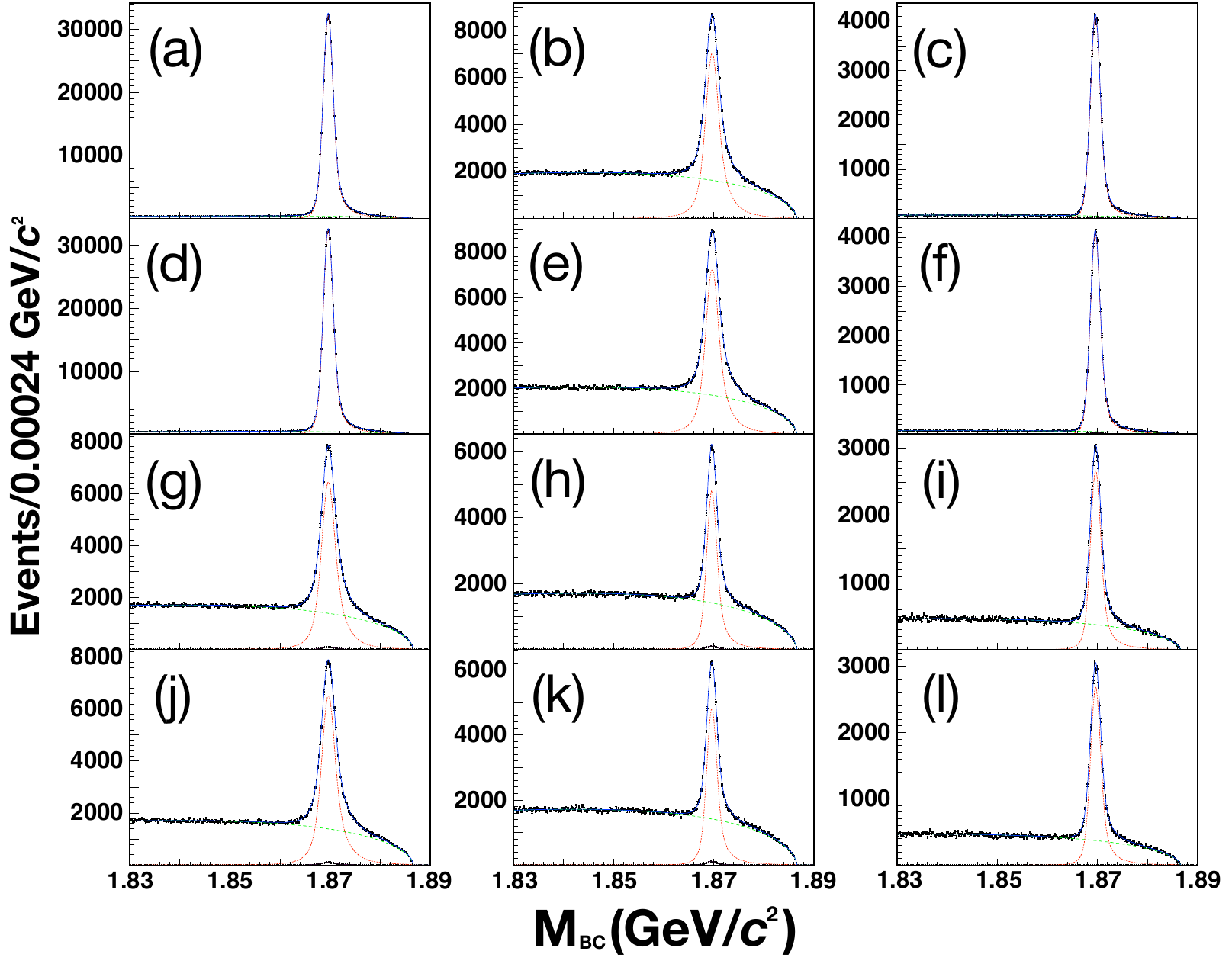}
	\figcaption{\label{stdp} $\mbc$ fits for single-tag modes;
		(a) $D^+\to K^-\pi^+\pi^+$, (b) $D^+\to K^-\pi^+\pi^+\pi^0$, (c) $D^+\to K_S^0\pi^+$,
		(d) $D^-\to K^+\pi^-\pi^-$, (e) $D^-\to K^+\pi^-\pi^-\pi^0$, (f) $D^-\to K_S^0\pi^-$,
		(g) $D^+\to K_S^0\pi^+\pi^0$, (h) $D^+\to K_S^0\pi^+\pi^+\pi^-$, (i) $D^+\to K^+K^-\pi^+$,
		(j) $D^-\to K_S^0\pi^-\pi^0$, (k) $D^-\to K_S^0\pi^+\pi^-\pi^-$, (l) $D^-\to K^+K^-\pi^-$.
		Blue solid, red dotted, and green dashed  lines represent the total fits,
		the fitted signal shapes, and the fitted background shapes, respectively,
		while black histograms correspond to the expected peaking background components. }
\end{center}

\begin{center}
	\includegraphics[width=13.5cm]{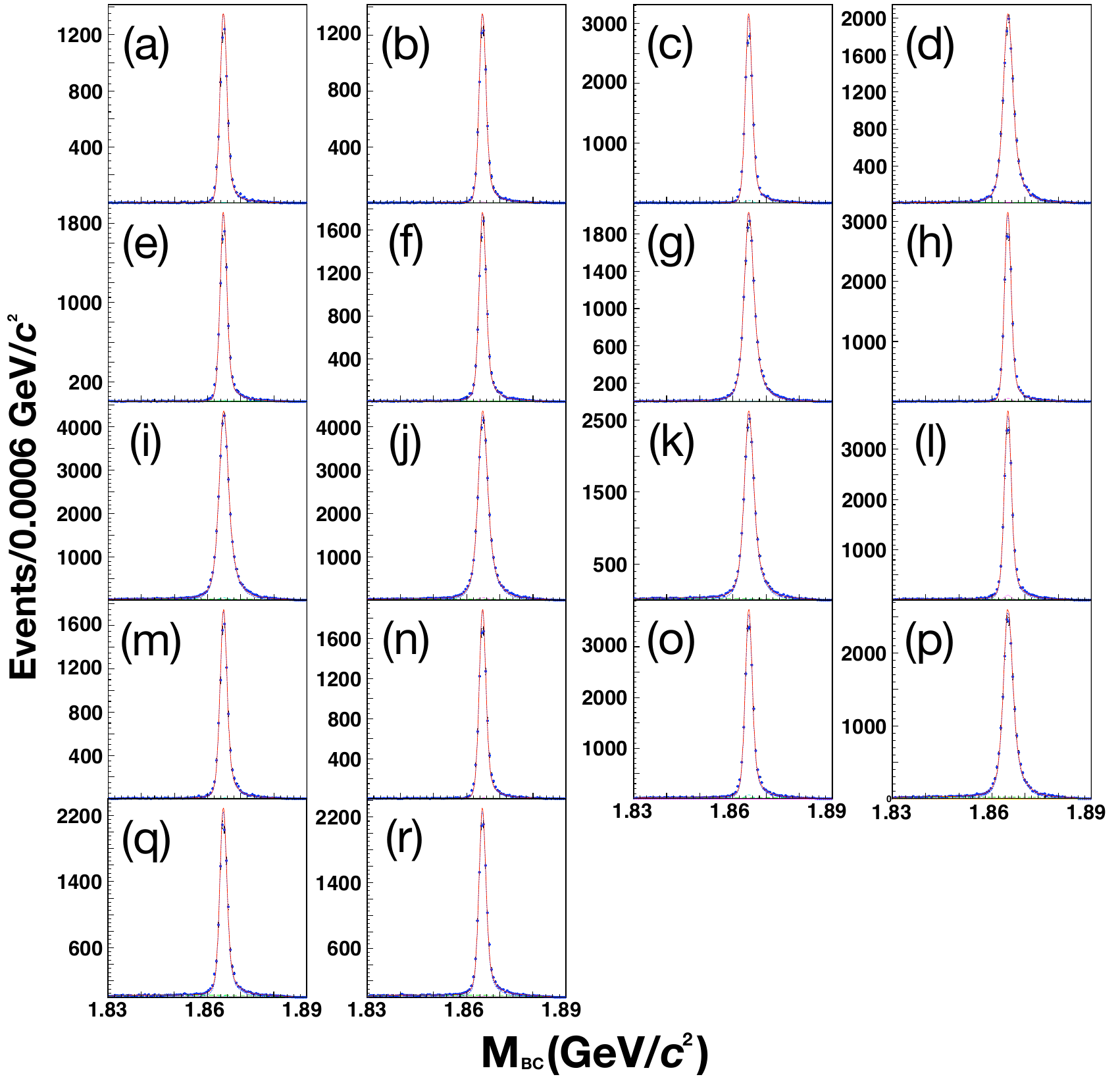}
	\figcaption{\label{dtd0} Two-dimensional $\mbc$ fits projected onto the
		positive and negative charm $\mbc$ axes for various double-tag modes;
		(a) $D^0\to K^-\pi^+$ vs. (b) $\bar{D}^0\to K^+\pi^-$,
		(c) $D^0\to K^-\pi^+$ vs. (d) $\bar{D}^0\to K^+\pi^-\pi^0$,
		(e) $D^0\to K^-\pi^+$ vs. (f) $\bar{D}^0\to K^+\pi^+\pi^-\pi^-$,
		(g) $D^0\to K^-\pi^+\pi^0$ vs. (h) $\bar{D}^0\to K^+\pi^-$,
		(i) $D^0\to K^-\pi^+\pi^0$ vs. (j) $\bar{D}^0\to K^+\pi^-\pi^0$,
		(k) $D^0\to K^-\pi^+\pi^0$ vs. (l) $\bar{D}^0\to K^+\pi^+\pi^-\pi^-$,
		(m) $D^0\to K^-\pi^+\pi^+\pi^-$ vs. (n) $\bar{D}^0\to K^+\pi^-$,
		(o) $D^0\to K^-\pi^+\pi^+\pi^-$ vs. (p) $\bar{D}^0\to K^+\pi^-\pi^0$,
		(q) $D^0\to K^-\pi^+\pi^+\pi^-$ vs. (r) $\bar{D}^0\to K^+\pi^+\pi^-\pi^-$.
		Red solid and blue dotted curves represent the total fits and
		the fitted signal shapes, respectively.
		Green long-dashed and orange solid lines correspond to
		the fitted non-peaking background shapes, while
		cyan and magenta short-dashed curves are the fitted peaking background components.}
\end{center}

\begin{center}
	\includegraphics[width=13.5cm]{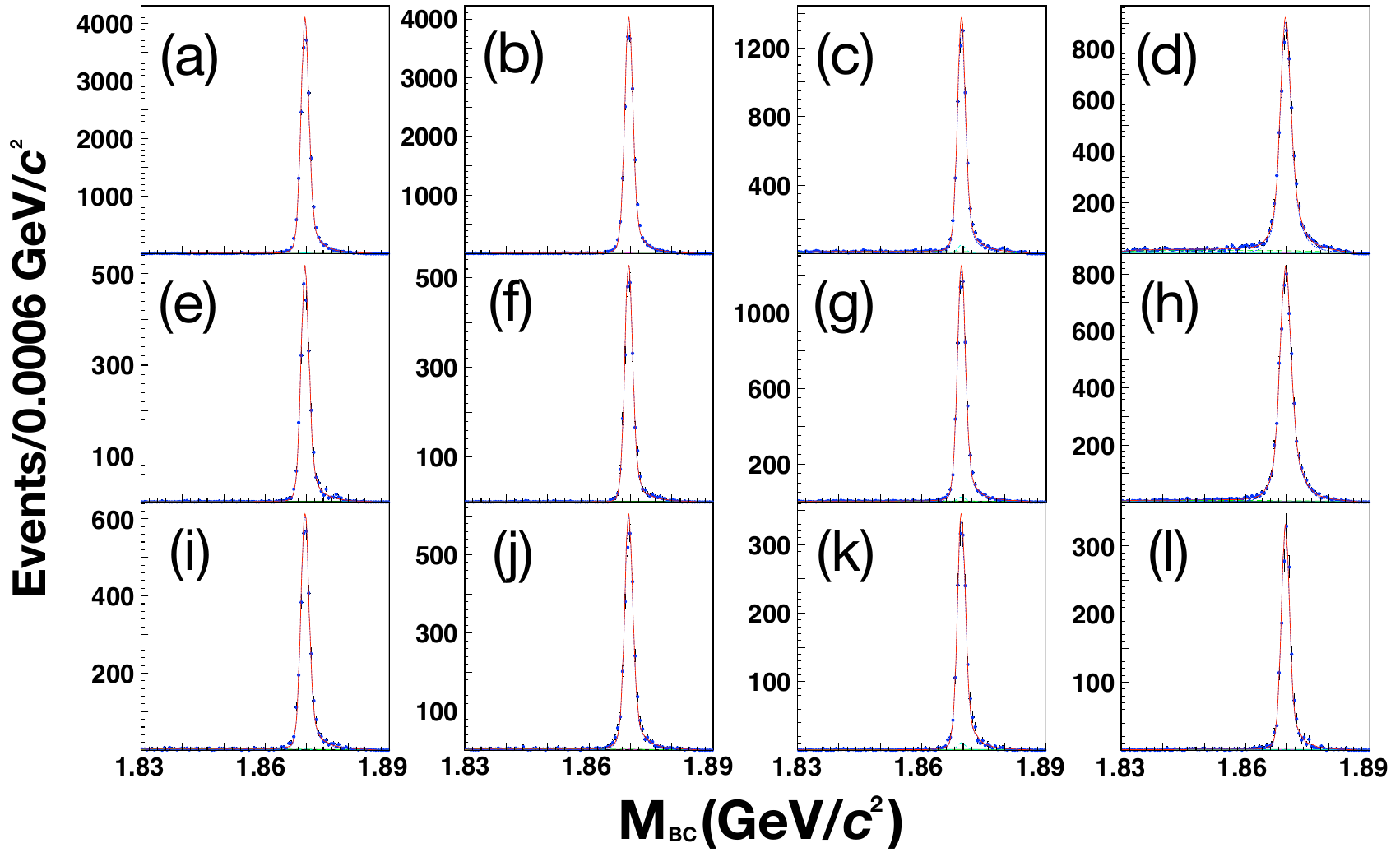}
	\figcaption{\label{dtdp200} Two-dimensional $\mbc$ fits projected onto the
		positive and negative charm $\mbc$ axes for various double-tag modes;
		(a) $D^+\to K^-\pi^+\pi^+$ vs. (b) $D^-\to K^+\pi^-\pi^-$,
		(c) $D^+\to K^-\pi^+\pi^+$ vs. (d) $D^-\to K^+\pi^-\pi^-\pi^0$,
		(e) $D^+\to K^-\pi^+\pi^+$ vs. (f) $D^-\to K_S^0\pi^-$,
		(g) $D^+\to K^-\pi^+\pi^+$ vs. (h) $D^-\to K_S^0\pi^-\pi^0$,
		(i) $D^+\to K^-\pi^+\pi^+$ vs. (j) $D^-\to K_S^0\pi^+\pi^-\pi^-$,
		(k) $D^+\to K^-\pi^+\pi^+$ vs. (l) $D^-\to K^+K^-\pi^-$.
		Red solid and blue dotted curves represent the total fits and
		the fitted signal shapes, respectively.
		Green long-dashed and orange solid lines correspond to
		the fitted non-peaking background shapes, while
		cyan and magenta short-dashed curves are the fitted peaking background components.}
\end{center}
\vspace{2mm}
\begin{center}
	\includegraphics[width=13.5cm]{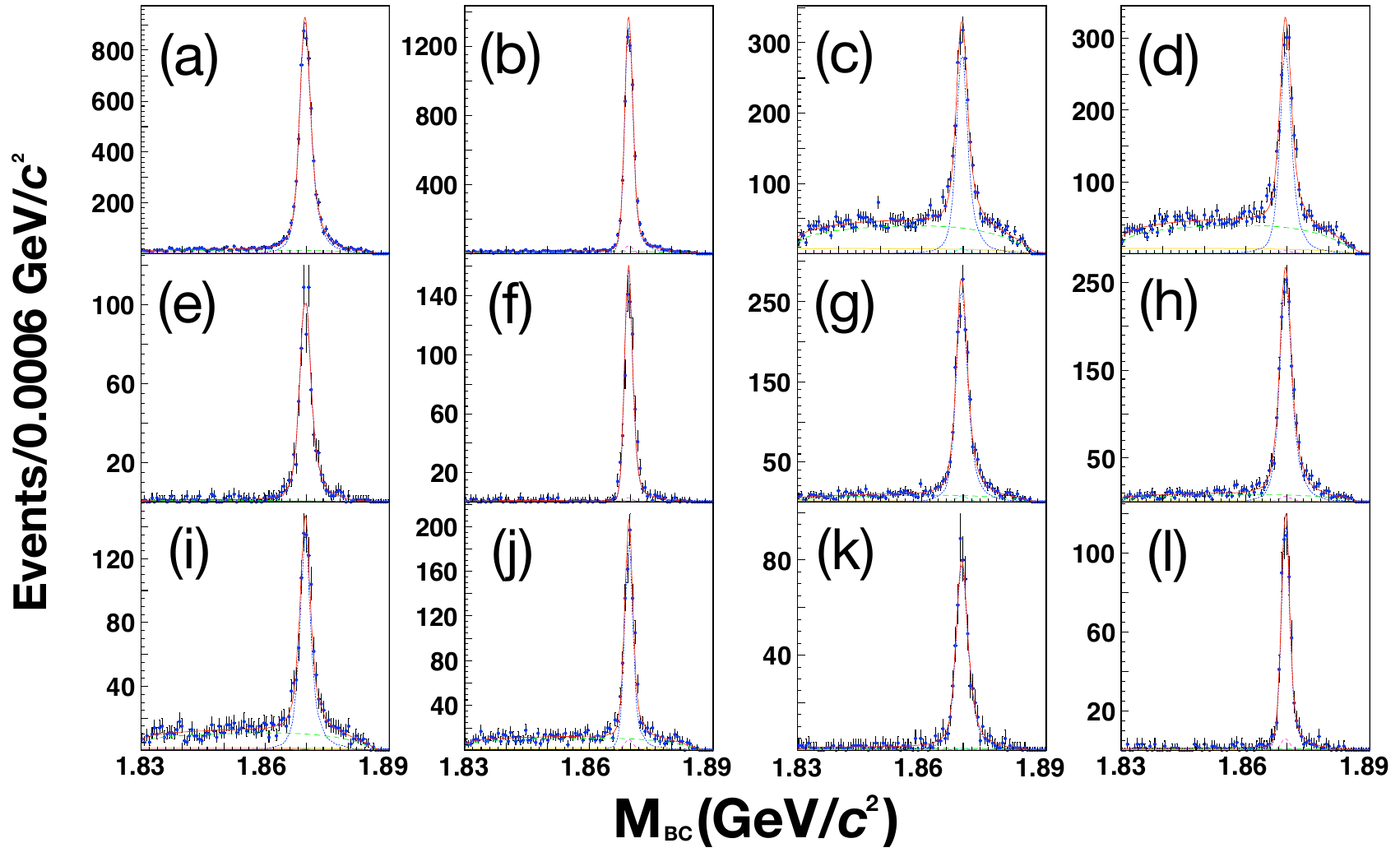}
	\figcaption{\label{dtdp201} Two-dimensional $\mbc$ fits projected onto the
		positive and negative charm $\mbc$ axes for various double-tag modes;
		(a) $D^+\to K^-\pi^+\pi^+\pi^0$ vs. (b) $D^-\to K^+\pi^-\pi^-$,
		(c) $D^+\to K^-\pi^+\pi^+\pi^0$ vs. (d) $D^-\to K^+\pi^-\pi^-\pi^0$,
		(e) $D^+\to K^-\pi^+\pi^+\pi^0$ vs. (f) $D^-\to K_S^0\pi^-$,
		(g) $D^+\to K^-\pi^+\pi^+\pi^0$ vs. (h) $D^-\to K_S^0\pi^-\pi^0$,
		(i) $D^+\to K^-\pi^+\pi^+\pi^0$ vs. (j) $D^-\to K_S^0\pi^+\pi^-\pi^-$,
		(k) $D^+\to K^-\pi^+\pi^+\pi^0$ vs. (l) $D^-\to K^+K^-\pi^-$.
		Red solid and blue dotted curves represent the total fits and
		the fitted signal shapes, respectively.
		Green long-dashed and orange solid lines correspond to
		the fitted non-peaking background shapes, while
		cyan and magenta short-dashed curves are the fitted peaking background components.}
\end{center}

\begin{center}
	\includegraphics[width=13.5cm]{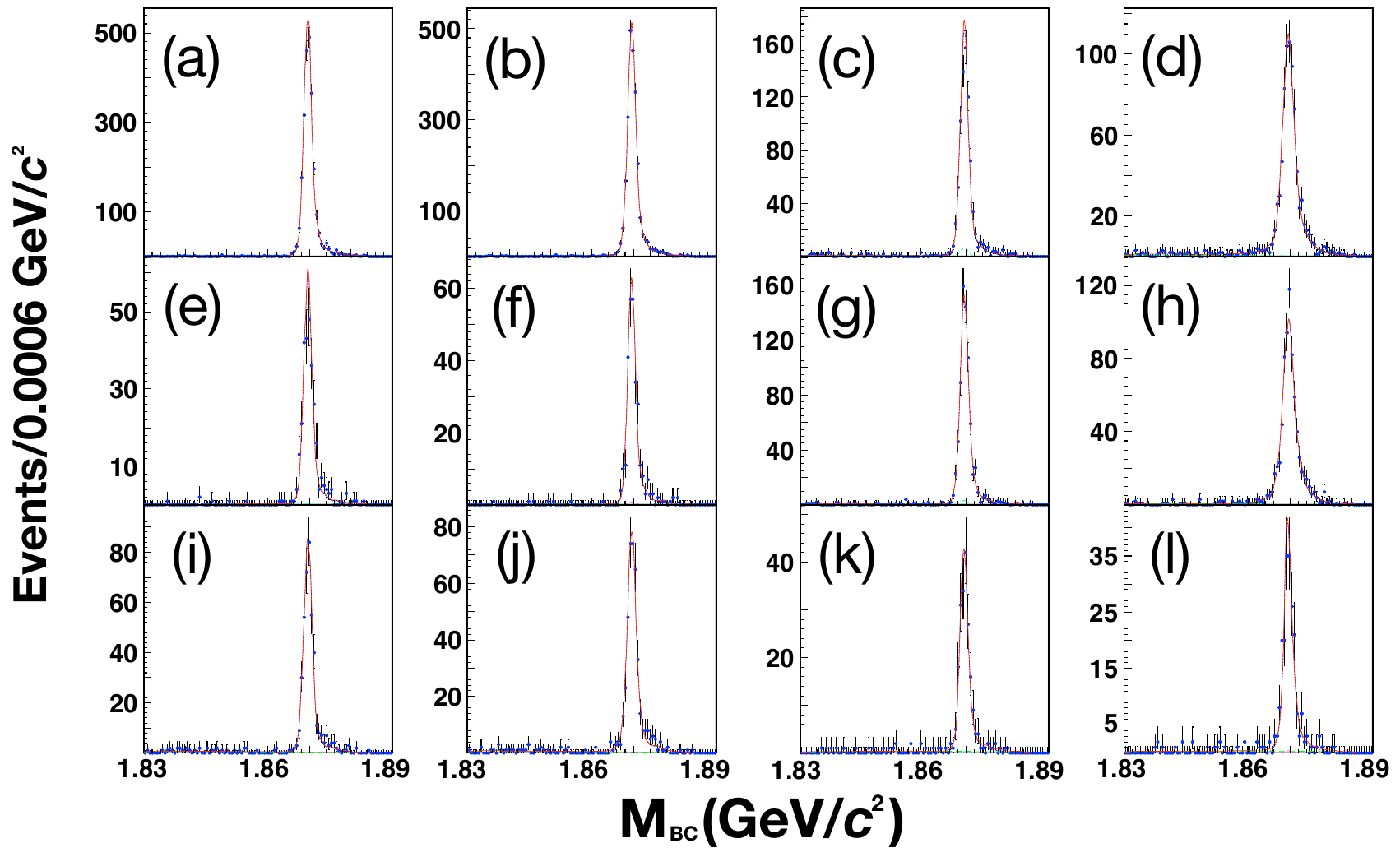}
	\figcaption{\label{dtdp202} Two-dimensional $\mbc$ fits projected onto the
		positive and negative charm $\mbc$ axes for various double-tag modes;
		(a) $D^+\to K_S^0\pi^+$ vs. (b) $D^-\to K^+\pi^-\pi^-$,
		(c) $D^+\to K_S^0\pi^+$ vs. (d) $D^-\to K^+\pi^-\pi^-\pi^0$,
		(e) $D^+\to K_S^0\pi^+$ vs. (f) $D^-\to K_S^0\pi^-$,
		(g) $D^+\to K_S^0\pi^+$ vs. (h) $D^-\to K_S^0\pi^-\pi^0$,
		(i) $D^+\to K_S^0\pi^+$ vs. (j) $D^-\to K_S^0\pi^+\pi^-\pi^-$,
		(k) $D^+\to K_S^0\pi^+$ vs. (l) $D^-\to K^+K^-\pi^-$.
		Red solid and blue dotted curves represent the total fits and
		the fitted signal shapes, respectively.
		Green long-dashed and orange solid lines correspond to
		the fitted non-peaking background shapes, while
		cyan and magenta short-dashed curves are the fitted peaking background components.}
\end{center}
\vspace{2mm}
\begin{center}
	\includegraphics[width=13.5cm]{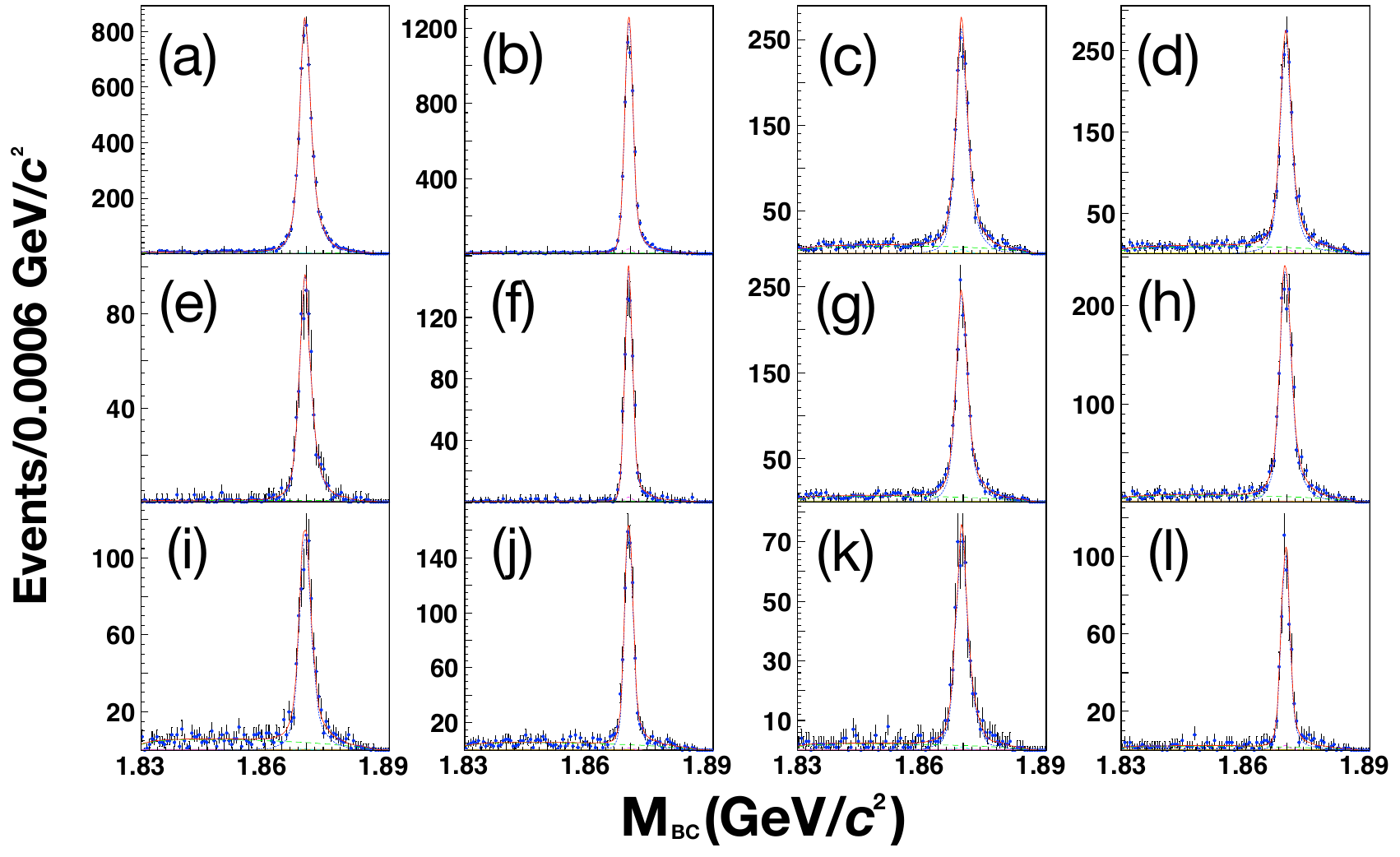}
	\figcaption{\label{dtdp203} Two-dimensional $\mbc$ fits projected onto the
		positive and negative charm $\mbc$ axes for various double-tag modes;
		(a) $D^+\to K_S^0\pi^+\pi^0$ vs. (b) $D^-\to K^+\pi^-\pi^-$,
		(c) $D^+\to K_S^0\pi^+\pi^0$ vs. (d) $D^-\to K^+\pi^-\pi^-\pi^0$,
		(e) $D^+\to K_S^0\pi^+\pi^0$ vs. (f) $D^-\to K_S^0\pi^-$,
		(g) $D^+\to K_S^0\pi^+\pi^0$ vs. (h) $D^-\to K_S^0\pi^-\pi^0$,
		(i) $D^+\to K_S^0\pi^+\pi^0$ vs. (j) $D^-\to K_S^0\pi^+\pi^-\pi^-$,
		(k) $D^+\to K_S^0\pi^+\pi^0$ vs. (l) $D^-\to K^+K^-\pi^-$.
		Red solid and blue dotted curves represent the total fits and
		the fitted signal shapes, respectively.
		Green long-dashed and orange solid lines correspond to
		the fitted non-peaking background shapes, while
		cyan and magenta short-dashed curves are the fitted peaking background components.}
\end{center}

\begin{center}
	\includegraphics[width=13.5cm]{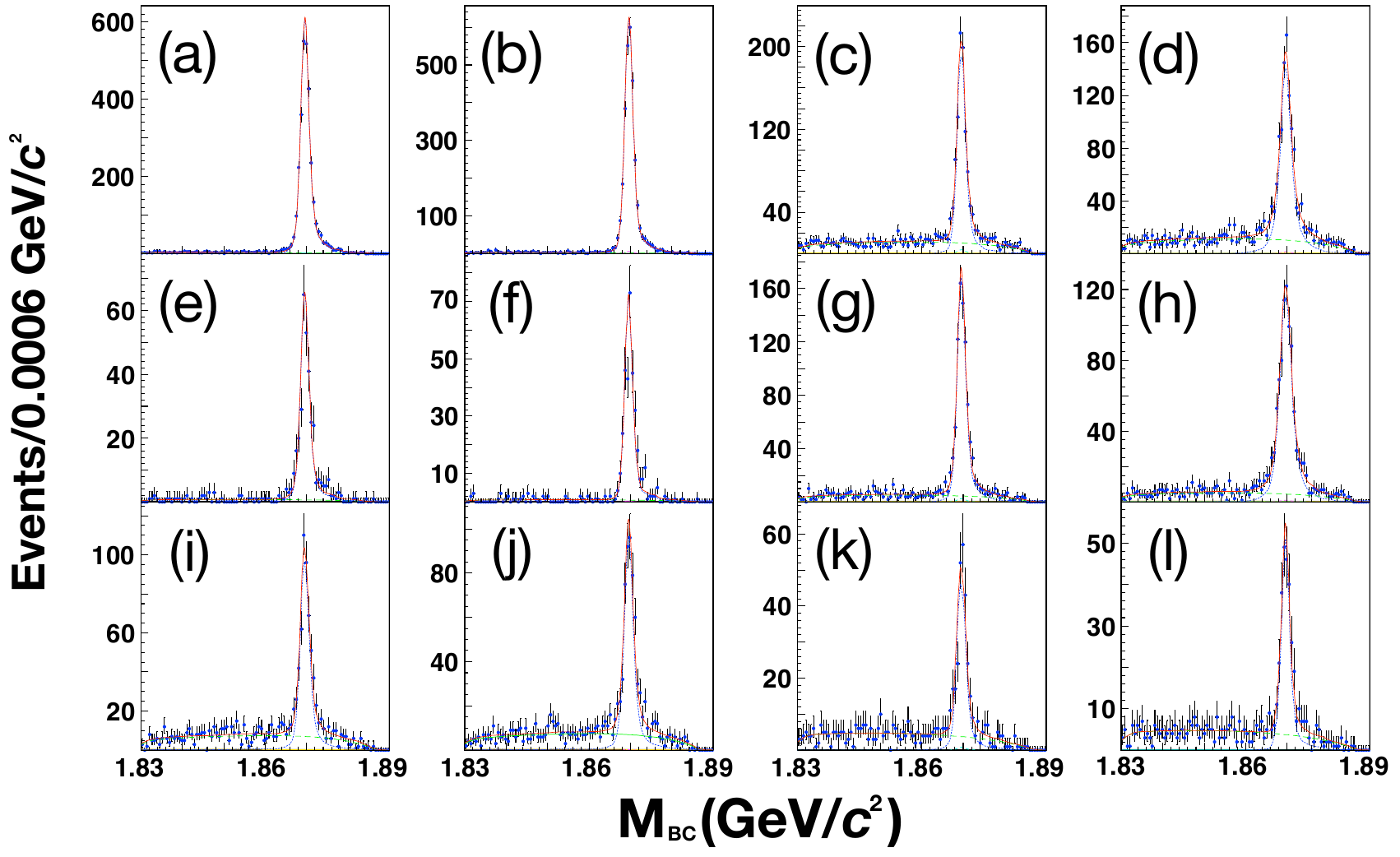}
	\figcaption{\label{dtdp204} Two-dimensional $\mbc$ fits projected onto the
		positive and negative charm $\mbc$ axes for various double-tag modes;
		(a) $D^+\to K_S^0\pi^+\pi^+\pi^-$ vs. (b) $D^-\to K^+\pi^-\pi^-$,
		(c) $D^+\to K_S^0\pi^+\pi^+\pi^-$ vs. (d) $D^-\to K^+\pi^-\pi^-\pi^0$,
		(e) $D^+\to K_S^0\pi^+\pi^+\pi^-$ vs. (f) $D^-\to K_S^0\pi^-$,
		(g) $D^+\to K_S^0\pi^+\pi^+\pi^-$ vs. (h) $D^-\to K_S^0\pi^-\pi^0$,
		(i) $D^+\to K_S^0\pi^+\pi^+\pi^-$ vs. (j) $D^-\to K_S^0\pi^+\pi^-\pi^-$,
		(k) $D^+\to K_S^0\pi^+\pi^+\pi^-$ vs. (l) $D^-\to K^+K^-\pi^-$.
		Red solid and blue dotted curves represent the total fits and
		the fitted signal shapes, respectively.
		Green long-dashed and orange solid lines correspond to
		the fitted non-peaking background shapes, while
		cyan and magenta short-dashed curves are the fitted peaking background components.}
\end{center}
\vspace{2mm}
\begin{center}
	\includegraphics[width=13.5cm]{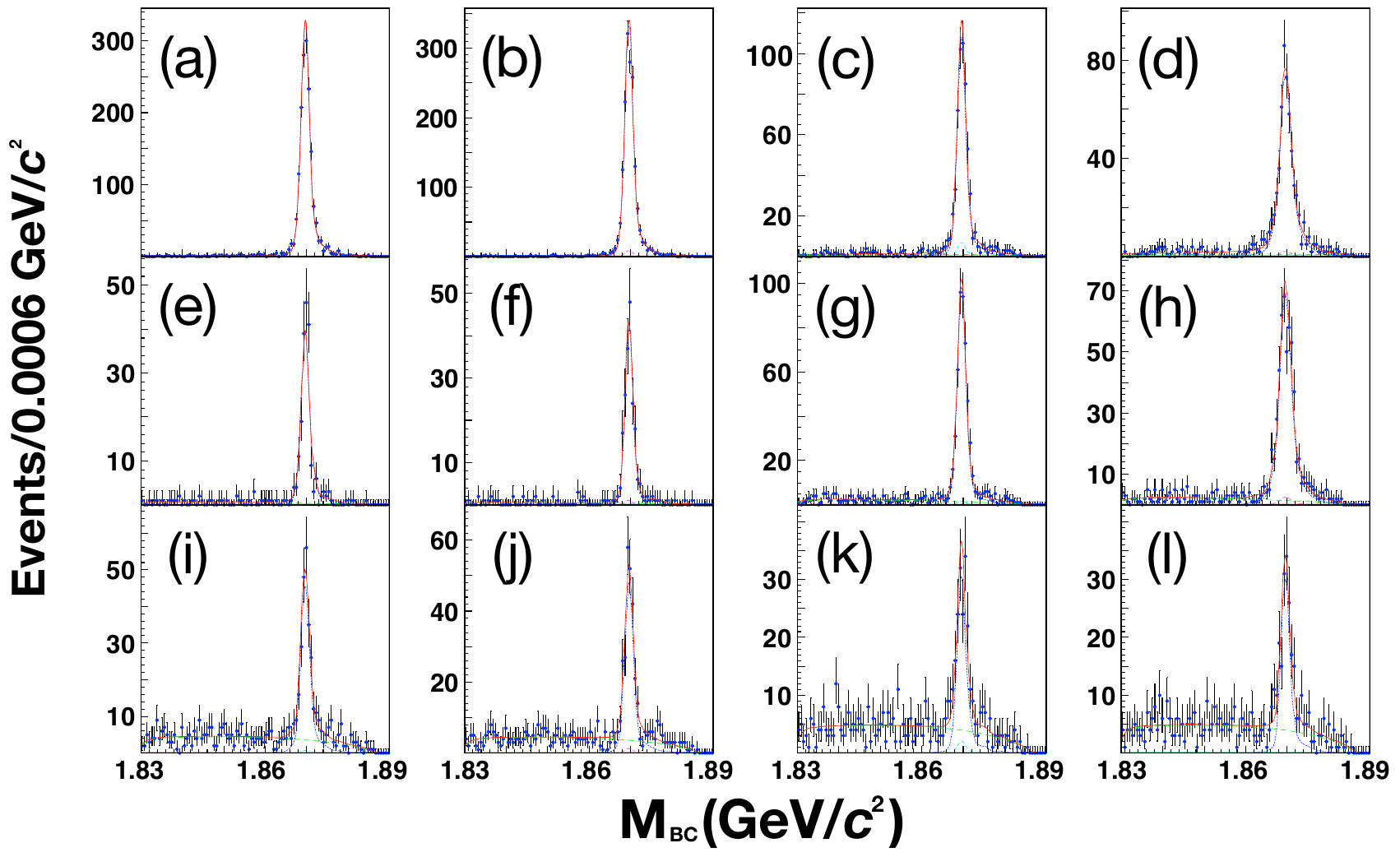}
	\figcaption{\label{dtdp205} Two-dimensional $\mbc$ fits projected onto the
		positive and negative charm $\mbc$ axes for various double-tag modes;
		(a) $D^+\to K^+K^-\pi^+$ vs. (b) $D^-\to K^+\pi^-\pi^-$,
		(c) $D^+\to K^+K^-\pi^+$ vs. (d) $D^-\to K^+\pi^-\pi^-\pi^0$,
		(e) $D^+\to K^+K^-\pi^+$ vs. (f) $D^-\to K_S^0\pi^-$,
		(g) $D^+\to K^+K^-\pi^+$ vs. (h) $D^-\to K_S^0\pi^-\pi^0$,
		(i) $D^+\to K^+K^-\pi^+$ vs. (j) $D^-\to K_S^0\pi^+\pi^-\pi^-$,
		(k) $D^+\to K^+K^-\pi^+$ vs. (l) $D^-\to K^+K^-\pi^-$.
		Red solid and blue dotted curves represent the total fits and
		the fitted signal shapes, respectively.
		Green long-dashed and orange solid lines correspond to
		the fitted non-peaking background shapes, while
		cyan and magenta short-dashed curves are the fitted peaking background components.}
\end{center}

\vspace{10mm}

\vspace{-1mm}
\centerline{\rule{80mm}{0.1pt}}
\vspace{2mm}

\begin{multicols}{2}

\end{multicols*}

\clearpage

\end{CJK*}
\end{document}